\catcode`\@=11					% To make protected \def's

%************************************************************
%*
%*		Font set-up
%*
%************************************************************

%************** 5-point fonts *******************************

\font\fiverm=cmr5				% roman
\font\fivemi=cmmi5				% math italic
\font\fivesy=cmsy5				% math symbols
\font\fivebf=cmbx5				% bold face

\skewchar\fivemi='177
\skewchar\fivesy='60

%************** 6-point fonts *******************************

\font\sixrm=cmr6				% roman
\font\sixi=cmmi6				% math italic
\font\sixsy=cmsy6				% math symbols
\font\sixbf=cmbx6				% bold face

\skewchar\sixi='177
\skewchar\sixsy='60

%************** 7-point fonts *******************************

\font\sevenrm=cmr7				% roman
\font\seveni=cmmi7				% math italic
\font\sevensy=cmsy7				% math symbols
\font\sevenit=cmti7				% italic
\font\sevenbf=cmbx7				% bold face

\skewchar\seveni='177
\skewchar\sevensy='60

%************** 8-point fonts *******************************

\font\eightrm=cmr8				% roman
\font\eighti=cmmi8				% math italic
\font\eightsy=cmsy8				% math symbols
\font\eightit=cmti8				% italic
				% slanted
\font\eightbf=cmbx8				% bold face
				% typewriter
				% sans serif

\skewchar\eighti='177
\skewchar\eightsy='60

%************** 9-point fonts *******************************

\font\ninei=cmmi9
\font\ninesy=cmsy9

\skewchar\ninei='177
\skewchar\ninesy='60

%************** 10-point fonts ******************************

\font\tenrm=cmr10				% roman
\font\teni=cmmi10				% math italic
\font\tensy=cmsy10				% math symbols
\font\tenex=cmex10				% math extension
\font\tenit=cmti10				% italic
\font\tensl=cmsl10				% slanted
\font\tenbf=cmbx10				% bold face
\font\tentt=cmtt10				% typewriter
\font\tenss=cmss10				% sans serif
\font\tensc=cmcsc10				% small caps
\font\tenbi=cmmib10				% bold math

\skewchar\teni='177
\skewchar\tenbi='177
\skewchar\tensy='60

\def\tenpoint{\ifmmode\err@badsizechange\else
	\textfont0=\tenrm \scriptfont0=\sevenrm \scriptscriptfont0=\fiverm
	\textfont1=\teni  \scriptfont1=\seveni  \scriptscriptfont1=\fivemi
	\textfont2=\tensy \scriptfont2=\sevensy \scriptscriptfont2=\fivesy
	\textfont3=\tenex \scriptfont3=\tenex   \scriptscriptfont3=\tenex
	\textfont4=\tenit \scriptfont4=\sevenit \scriptscriptfont4=\sevenit
	\textfont5=\tensl
	\textfont6=\tenbf \scriptfont6=\sevenbf \scriptscriptfont6=\fivebf
	\textfont7=\tentt
	\textfont8=\tenbi \scriptfont8=\seveni  \scriptscriptfont8=\fivemi
	\def\rm{\tenrm\fam=0 }%
	\def\it{\tenit\fam=4 }%
	\def\sl{\tensl\fam=5 }%
	\def\bf{\tenbf\fam=6 }%
	\def\tt{\tentt\fam=7 }%
	\def\ss{\tenss}%
	\def\sc{\tensc}%
	\def\bmit{\fam=8 }%
	\rm\setparameters\setbaselines\fi}

%************** 12-point fonts ******************************

\font\twelverm=cmr12				% roman
\font\twelvei=cmmi12				% math italic
\font\twelvesy=cmsy10	scaled\magstep1		% math symbols
\font\twelveex=cmex10	scaled\magstep1		% math extension
\font\twelveit=cmti12				% italic
\font\twelvesl=cmsl12				% slanted
\font\twelvebf=cmbx12				% bold face
\font\twelvett=cmtt12				% typewriter
\font\twelvess=cmss12				% sans serif
\font\twelvesc=cmcsc10	scaled\magstep1		% small caps
\font\twelvebi=cmmib10	scaled\magstep1		% bold math

\skewchar\twelvei='177
\skewchar\twelvebi='177
\skewchar\twelvesy='60

\def\twelvepoint{\ifmmode\err@badsizechange\else
	\textfont0=\twelverm \scriptfont0=\eightrm \scriptscriptfont0=\sixrm
	\textfont1=\twelvei  \scriptfont1=\eighti  \scriptscriptfont1=\sixi
	\textfont2=\twelvesy \scriptfont2=\eightsy \scriptscriptfont2=\sixsy
	\textfont3=\twelveex \scriptfont3=\tenex   \scriptscriptfont3=\tenex
	\textfont4=\twelveit \scriptfont4=\eightit \scriptscriptfont4=\sevenit
	\textfont5=\twelvesl
	\textfont6=\twelvebf \scriptfont6=\eightbf \scriptscriptfont6=\sixbf
	\textfont7=\twelvett
	\textfont8=\twelvebi \scriptfont8=\eighti  \scriptscriptfont8=\sixi
	\def\rm{\twelverm\fam=0 }%
	\def\it{\twelveit\fam=4 }%
	\def\sl{\twelvesl\fam=5 }%
	\def\bf{\twelvebf\fam=6 }%
	\def\tt{\twelvett\fam=7 }%
	\def\ss{\twelvess}%
	\def\sc{\twelvesc}%
	\def\bmit{\fam=8 }%
	\rm\setparameters\setbaselines\fi}

%************** 14-point fonts ******************************

\font\fourteenrm=cmr12	scaled\magstep1		% roman
\font\fourteeni=cmmi12	scaled\magstep1		% math italic
\font\fourteensy=cmsy10	scaled\magstep2		% math symbols
\font\fourteenex=cmex10	scaled\magstep2		% math extension
\font\fourteenit=cmti12	scaled\magstep1		% italic
\font\fourteensl=cmsl12	scaled\magstep1		% slanted
\font\fourteenbf=cmbx12	scaled\magstep1		% bold face
\font\fourteentt=cmtt12	scaled\magstep1		% typewriter
\font\fourteenss=cmss12	scaled\magstep1		% sans serif
\font\fourteensc=cmcsc10 scaled\magstep2	% small caps
\font\fourteenbi=cmmib10 scaled\magstep2	% bold math

\skewchar\fourteeni='177
\skewchar\fourteenbi='177
\skewchar\fourteensy='60

\def\fourteenpoint{\ifmmode\err@badsizechange\else
	\textfont0=\fourteenrm \scriptfont0=\tenrm \scriptscriptfont0=\sevenrm
	\textfont1=\fourteeni  \scriptfont1=\teni  \scriptscriptfont1=\seveni
	\textfont2=\fourteensy \scriptfont2=\tensy \scriptscriptfont2=\sevensy
	\textfont3=\fourteenex \scriptfont3=\tenex \scriptscriptfont3=\tenex
	\textfont4=\fourteenit \scriptfont4=\tenit \scriptscriptfont4=\sevenit
	\textfont5=\fourteensl
	\textfont6=\fourteenbf \scriptfont6=\tenbf \scriptscriptfont6=\sevenbf
	\textfont7=\fourteentt
	\textfont8=\fourteenbi \scriptfont8=\tenbi \scriptscriptfont8=\seveni
	\def\rm{\fourteenrm\fam=0 }%
	\def\it{\fourteenit\fam=4 }%
	\def\sl{\fourteensl\fam=5 }%
	\def\bf{\fourteenbf\fam=6 }%
	\def\tt{\fourteentt\fam=7}%
	\def\ss{\fourteenss}%
	\def\sc{\fourteensc}%
	\def\bmit{\fam=8 }%
	\rm\setparameters\setbaselines\fi}

%************** Miscellaneous big fonts *********************

\font\seventeenrm=cmr10 scaled\magstep3		% roman
		% bold face

%************************************************************
%*
%*		Parameter initialization
%*
%************************************************************

\newdimen\rp@
\newcount\@basestretchnum
\newskip\@baseskip
\newskip\headskip
\newskip\footskip

% Routine to set page parameters

\def\setparameters{\rp@=.1em
	\headskip=24\rp@
	\footskip=\headskip
	\delimitershortfall=5\rp@
	\nulldelimiterspace=1.2\rp@
	\scriptspace=0.5\rp@
	\abovedisplayskip=10\rp@ plus3\rp@ minus5\rp@
	\belowdisplayskip=10\rp@ plus3\rp@ minus5\rp@
	\abovedisplayshortskip=5\rp@ plus2\rp@ minus4\rp@
	\belowdisplayshortskip=10\rp@ plus3\rp@ minus5\rp@
	\normallineskip=\rp@
	\lineskip=\normallineskip
	\normallineskiplimit=0pt
	\lineskiplimit=\normallineskiplimit
	\jot=3\rp@
	\setbox0=\hbox{\the\textfont3 B}\p@renwd=\wd0
	\skip\footins=12\rp@ plus3\rp@ minus3\rp@
	\skip\topins=0pt plus0pt minus0pt}

% Special routine to scale \baselineskip

\def\setbaselines{\maxdepth=4\rp@\baselinestretch=\@basestretchnum}

% The \baselinestretch command

\def\baselinestretch{\afterassignment\@basestretch\@basestretchnum}
\def\@basestretch{%
	\@baseskip=12\rp@ \divide\@baseskip by1000
	\normalbaselineskip=\@basestretchnum\@baseskip
	\baselineskip=\normalbaselineskip
	\bigskipamount=\the\baselineskip
		plus.25\baselineskip minus.25\baselineskip
	\medskipamount=.5\baselineskip
		plus.125\baselineskip minus.125\baselineskip
	\smallskipamount=.25\baselineskip
		plus.0625\baselineskip minus.0625\baselineskip
	\setbox\strutbox=\hbox{\vrule height.708\baselineskip
		depth.292\baselineskip width0pt }}

%************************************************************
%*
%*		Modifications to PLAIN.TEX
%*
%************************************************************

% Modifications to PLAIN routines to handle scaling of page parameters

\def\makeheadline{\vbox to0pt{\baselinestretch=1000
	\vskip-\headskip \vskip1.5pt
	\line{\vbox to\ht\strutbox{}\the\headline}\vss}\nointerlineskip}

\def\makefootline{\baselineskip=\footskip\line{\the\footline}}

\def\big#1{{\hbox{$\left#1\vbox to8.5\rp@ {}\right.\n@space$}}}
\def\Big#1{{\hbox{$\left#1\vbox to11.5\rp@ {}\right.\n@space$}}}
\def\bigg#1{{\hbox{$\left#1\vbox to14.5\rp@ {}\right.\n@space$}}}
\def\Bigg#1{{\hbox{$\left#1\vbox to17.5\rp@ {}\right.\n@space$}}}

% Modifications to PLAIN to handle bold math

\mathchardef\alpha="710B
\mathchardef\beta="710C
\mathchardef\gamma="710D
\mathchardef\delta="710E
\mathchardef\epsilon="710F
\mathchardef\zeta="7110
\mathchardef\eta="7111
\mathchardef\theta="7112
\mathchardef\iota="7113
\mathchardef\kappa="7114
\mathchardef\lambda="7115
\mathchardef\mu="7116
\mathchardef\nu="7117
\mathchardef\xi="7118
\mathchardef\pi="7119
\mathchardef\rho="711A
\mathchardef\sigma="711B
\mathchardef\tau="711C
\mathchardef\upsilon="711D
\mathchardef\phi="711E
\mathchardef\chi="711F
\mathchardef\psi="7120
\mathchardef\omega="7121
\mathchardef\varepsilon="7122
\mathchardef\vartheta="7123
\mathchardef\varpi="7124
\mathchardef\varrho="7125
\mathchardef\varsigma="7126
\mathchardef\varphi="7127
\mathchardef\imath="717B
\mathchardef\jmath="717C
\mathchardef\ell="7160
\mathchardef\wp="717D
\mathchardef\partial="7140
\mathchardef\flat="715B
\mathchardef\natural="715C
\mathchardef\sharp="715D

%************************************************************
%*
%*		Initialization
%*
%************************************************************

\def\err@badsizechange{%
	\immediate\write16{--> Size change not allowed in math mode, ignored}}

\baselinestretch=1000
\tenpoint

\catcode`\@=12					% Restore @ sign
% Routine to guarantee that this file is input only once
\catcode`\@=11
\expandafter\ifx\csname @iasmacros\endcsname\relax
	\global\let\@iasmacros=\par
\else	\immediate\write16{}
	\immediate\write16{Warning:}
	\immediate\write16{You have tried to input iasmacros more than once.}
	\immediate\write16{}
	\endinput
\fi
\catcode`\@=12

% Set up font size commands and \baselinestretch command
%\input iasfonts

% Some alternative font names
\def\rmb{\seventeenrm}

% Simple spacing commands
\def\singlespace{\baselineskip=\normalbaselineskip}
\def\halfspace{\baselineskip=1.5\normalbaselineskip}
\def\doublespace{\baselineskip=2\normalbaselineskip}

% Macros for references and abstracts

\def\AB{\bigskip\parindent=40pt
        \centerline{\bf ABSTRACT}\medskip\halfspace\narrower}
\def\AE{\bigskip\nonarrower\doublespace}
\def\nonarrower{\advance\leftskip by-\parindent
	\advance\rightskip by-\parindent}

% Useful commands

\def\boxit#1{\vbox{\hrule\hbox{\vrule\kern3pt
	\vbox{\kern3pt#1\kern3pt}\kern3pt\vrule}\hrule}}

% Special symbols
\def\hence{\leavevmode\hbox{\bf .\raise5.5pt\hbox{.}.} }

\def\dalemb#1#2{{\vbox{\hrule height.#2pt
	\hbox{\vrule width.#2pt height#1pt \kern#1pt \vrule width.#2pt}
	\hrule height.#2pt}}}
\def\gtorder{\mathrel{\raise.3ex\hbox{$>$}\mkern-14mu
             \lower0.6ex\hbox{$\sim$}}}
\def\ltorder{\mathrel{\raise.3ex\hbox{$<$}\mkern-14mu
             \lower0.6ex\hbox{$\sim$}}}

% For twoup output
\newdimen\fullhsize
\newbox\leftcolumn
\def\twoup{\hoffset=-.5in \voffset=-.25in
  \hsize=4.75in \fullhsize=10in \vsize=6.9in
  \def\fullline{\hbox to\fullhsize}
  \let\lr=L
  \output={\if L\lr
        \global\setbox\leftcolumn=\columnbox\global\let\lr=R \advancepageno
      \else \doubleformat \global\let\lr=L\fi
    \ifnum\outputpenalty>-20000 \else\dosupereject\fi}
  \def\doubleformat{\shipout\vbox{
    \fullline{\box\leftcolumn\hfil\columnbox}\advancepageno}}
  \def\columnbox{\leftline{\vbox{\makeheadline\pagebody\makefootline}}}
  \tolerance=1000 }

\twelvepoint
\doublespace
{\nopagenumbers{
\rightline{IASSNS-HEP-97/125}
\rightline{~~~November, 1997}
\bigskip\bigskip
\centerline{\rmb A Model for the Quark Mass and Flavor Mixing Matrices Based} 
\centerline{\rmb on Discrete Chiral Symmetry as the Origin of Families}
\medskip
\centerline{\it Stephen L. Adler}
\centerline{\bf Institute for Advanced Study}
\centerline{\bf Princeton, NJ 08540}
\medskip
\bigskip\bigskip
\leftline{\it Send correspondence to:}
\medskip
{\singlespace\leftline{Stephen L. Adler}
\leftline{Institute for Advanced Study}
\leftline{Olden Lane, Princeton, NJ 08540}
\leftline{Phone 609-734-8051; FAX 609-924-8399; email adler@ias.edu}}
\bigskip\bigskip
}}
\vfill\eject
\pageno=2
\AB
We construct a model for the quark mass and flavor mixing matrices, based on 
the hypothesis that the flavor weak eigenstates in the three families are 
distinguished by a spontaneously broken discrete $Z_6$ chiral symmetry.
In a leading partially symmetric approximation, the model accommodates 
the family 
mass spectra, with a CKM matrix that is exactly unity.  Adding small 
asymmetries in first order perturbation theory gives a CKM matrix with 
the correct qualitative structure, with $s_{23}/s_{12}$ and $s_{13}/s_{12}$ 
naturally of order $M_s/M_b$.  
\AE
\bigskip\bigskip
\vfill\eject
\pageno=3
It has long been recognized that the hierarchical structures of the 
family mass spectra, with their large third family masses, and of the CKM 
mixing matrix, with its suppressed third family mixings, may have a common
dynamical origin.  In particular, several authors [1] have stressed 
that the 
observed pattern seems to be close to the ``rank-one'' limit, in which the 
mass matrices have the ``democratic'' form of a matrix with all matrix 
elements equal to unity, which has one eigenvalue 1 and two eigenvalues 0;  
when both up and down quark mass matrices have this form, 
they are diagonalized 
by the same unitary transformation and the CKM matrix is unity.  However, 
because the underlying dynamical basis for this choice has not been 
apparent, it has not been possible to systematically extend the rank-one   
model to one that incorporates, and relates, the observed mass and mixing 
hierarchies.  

We present in this letter a model for the quark mass and flavor mixing 
matrices, based on the underlying dynamical assumption that the three 
flavor weak eigenstates are distinguished by different eigenvalues of 
a discrete chiral $Z_6$ quantum number.  The idea that a discrete chiral 
quantum number may underlie family structure was introduced originally  
by Harari and Seiberg [2], and was developed recently by 
the author [3] in a
modified form that we follow here.  Also of relevance is the remark of  
Weinberg [4] that an unbroken discrete chiral quantum number suffices 
to enforce the masslessness of fermionic states.  Extending the general  
framework of this earlier work, we postulate that
all {\it complex} fields carry a discrete chiral family quantum number.  
Since the Higgs scalars in the standard model are complex, we introduce a 
triplet of Higgs doublets that carry $Z_6$ quantum numbers, and that 
are coupled to the fermions by Yukawa couplings 
so that the Lagrangian is $Z_6$ invariant.  
Spontaneous symmetry breaking, in which the neutral members of the three
Higgs doublets acquire vacuum expectations, then gives the 
fermion mass matrices that form the basis for our detailed analysis.  

In the simplest form of a theory with a discrete chiral $Z_6$ symmetry, 
the fundamental Lagrangian, as augmented by the instanton-induced potential, 
is invariant under the simultaneous transformations 
$$\xi_L \to \xi_L \exp(-2\pi i/6)~,~~~\xi_R \to \xi_R \exp(2\pi i/6)~~~
\eqno(1a)$$
of the unification scale fermion fields $\xi$.  
The fields in the low energy 
effective Lagrangian are in general nonlinear functionals of the fundamental 
fields.  Fermionic effective fields must be odd monomials in the fundamental 
fields, and so can come in three varieties $\psi_n$ with the discrete 
chiral transformation law
$$\psi_{nL} \to \psi_{nL} \exp(-(2n+1)2\pi i/6)~,~~~
\psi_{nR} \to \psi_{nR} \exp((2n+1)2\pi i/6)~,~~~
n=0,1,2~~~,\eqno(1b)$$
while complex bosonic effective fields must be even monomials in 
the fundamental fermion fields, and so can 
also come in three varieties $\phi_n$ 
with the discrete chiral transformation law 
$$\phi_n \to \phi_n \exp(2n 2\pi i/6)~,~~~n=0,1,2~~~.\eqno(1c)$$
Gauge fields are real fields, and since the phase in Eq.~(1c) never takes 
the value $-1$, they come in only one variety transforming with phase 
unity under discrete chiral transformations.   (Two 
varieties of gauge fields become possible when the discrete chiral group is 
$Z_{12}$ rather than $Z_6$, since $-1$ is then present in the set of even 
phases.)  Thus the minimal $Z_6$ invariant extension of the standard  
model consists of a triplicated set of fermions, and a triplicated set of 
Higgs doublets, obeying the transformation laws of Eqs.~(1b) and (1c) 
respectively, together with the usual gauge bosons,  with the Lagrangian 
constructed to be $Z_6$ invariant.  

We shall give the full Lagrangian 
structure of this theory elsewhere; 
for our present purpose it suffices to state 
that all of the usual gauge sector results are unchanged. Hence we only   
exhibit the most general Yukawa coupling of the Higgs fields to the fermions,
which grouping the the fermions $\psi_n$ into a 3-component column vector, 
takes the form 
$$\overline{Q}_L \Phi^d d_R + \overline{Q}_L \Phi^u u_R+
\overline{\psi}_L \Phi^e e_R~~~+ {\rm adjoint}.\eqno(2)$$
Here $Q_L$ and $\psi_L$ denote the usual left handed quark and lepton 
doublets, as extended into three component column vectors of doublets 
to incorporate the discrete chiral symmetry.  Similarly, $d_R$, $u_R$, 
and $e_R$ denote the usual quark and lepton right handed singlets, 
as extended 
into three component column vectors.  The $3 \times 3$ matrices $\Phi^{d,u,e}$ 

are given by 
$$\Phi^{d,e}=\sum_s P_s^{d,e} \phi_{n(s)}~,~~~
\Phi^{u}=\sum_s P_s^u \tilde \phi_{\tilde n(s)}~~~,\eqno(3a)$$
with $\phi$ and $\tilde \phi$ denoting respectively the Higgs doublet and 
the charge conjugate Higgs doublet, with $n(s)~,~ \tilde n(s)$ 
denoting the unique discrete 
chiral Higgs component, for each $s=1,2,3$, that cancels the fermionic phase 
selected by the combination matrices $P_s$. The three possibilities for $P_s$ 
are given by (with $f$ denoting $d,u,e$) 
$$P_0^{f}=\pmatrix{0       & 0       & \alpha_{13}^f    \cr
                   0       & \alpha_{22}^f   &0        \cr
                   \alpha_{31}^f &0  &0   \cr}~,~~~
P_1^{f}=\pmatrix{\alpha_{11}^f &0 &0 \cr
                 0 & 0& \alpha_{23}^f \cr
                 0 & \alpha_{32}^f & 0 \cr}~,~~~
P_2^{f}=\pmatrix{0 & \alpha_{12}^f & 0 \cr
                 \alpha_{21}^f & 0 & 0 \cr
                 0 & 0 & \alpha_{33}^f \cr}~~~,
\eqno(3b)$$
with the $\alpha$'s complex numbers.  On spontaneous symmetry breaking, 
the neutral components of each Higgs field $\phi_n$ obtains a vacuum 
expectation value $v_n$, and the Yukawa coupling of Eq.~(2) gives 
the mass term 
$$\sum_s[\overline{d}_L P_s^d v_{n(s)} d_R 
+ \overline{u}_L P_s^u v_{\tilde  n(s)} u_R
+\overline{e}_L P_s^e v_{n(s)} e_R~+ {\rm adjoint}].\eqno(4)$$
Equation (4) is the starting point for our analysis.  

We observe that 
there are two specializations of Eq.~(4) that are of interest.  The first 
is what we term the {\it partially symmetric} limit, in which the three Higgs 
expectations $v_n$ are not equal, but in which the $\alpha$'s appearing 
in each $P_s^f$ are equal to a common value $\alpha_s^f$, 
so that each Higgs couples universally to all three 
combinations of fermions that have the same overall phase under discrete 
chiral transformation.  When the partially symmetric limit is further 
specialized to the completely symmetric limit in which the products 
$\alpha_s^f v_{n(s)}$  (for $f=d,e$)  and $\alpha_s^f v_{\tilde n(s)}$ 
(for $f=u$) 
are independent of $s$, the mass matrices of Eq.~(4) for each charge 
sector become 
proportional to the ``democratic'' mass matrix with all matrix elements 
equal to unity.  

In the partially symmetric limit, the mass matrices are all proportional 
to a matrix of the general form 
$$N=\pmatrix{m_1 & m_2 & m_0\cr   
           m_2 & m_0 & m_1 \cr
           m_0 & m_1 & m_2\cr} ~~~.\eqno(5)$$
This matrix is a {\it circulant} [5] (the rows are related to one another 
by successive cyclic permutation), and thus has the property that it 
can be diagonalized by           
left and right unitary transformation matrices that are {\it independent} of 
the values of the complex numbers $m_0$, $m_1$, and $m_2$.  To show this, 
we introduce the unitary matrices $U_L$ and $U_R^{\dagger}$ given by 
$$U_L={1 \over \sqrt{3}}\pmatrix{1 & \omega^2 & \omega \cr
                                 1 & \omega & \omega^2 \cr
                                 1 & 1 & 1 \cr}~,~~~
U_R^{\dagger}={1 \over \sqrt{3}}\pmatrix{1 & 1 & 1 \cr
                               \omega^2 & \omega & 1 \cr
                               \omega & \omega^2 & 1 \cr}~~~,
                               \eqno(6a)$$
with $\omega$ the complex cube root of unity,                                
$$\eqalign{
\omega=&-{1\over 2}+{\sqrt{3}\over 2} i~,~~~
\omega^2=\omega^*= -{1\over 2}-{\sqrt{3}\over 2} i~~~~ \cr
\omega^3=&1~,~~~1+\omega+\omega^2=0~,~~~i(\omega^2-\omega)=\sqrt{3} ~~~.\cr
}\eqno(6b)$$
We then find that 
$$U_L N U_R^{\dagger}=\pmatrix{ n_1 & 0 & 0 \cr
                                0   & n_2 & 0 \cr
                                0 & 0 & n_3 \cr}~~~,\eqno(7a)$$
with the eigenvalues $n_{1,2,3}$ given by                                 
$$n_1=m_1+\omega^2 m_2+\omega m_0 ~,~~~
n_2= m_1+\omega m_2+\omega^2 m_0~,~~~
n_3=m_1+m_2+m_0~~~.\eqno(7b)$$
Because the transformation of Eq.~(6a) is independent of the values of 
the matrix elements of $N$, in the partially symmetric limit changing  
basis from weak eigenstates to mass eigenstates diagonalizes not only the 
mass term of Eq.~(4), but also the Yukawa coupling of Eq.~(2).  Hence  
in the partially symmetric limit there are no flavor changing neutral 
currents arising from tree level Higgs exchange.

To see that the magnitudes $|n_1|,|n_2|,|n_3|$ can represent arbitrary 
first, second, and third generation masses $M_1,M_2,M_3$, let us specialize 
to the case when $m_0$ is real and 
$m_2$ and $m_1$ are related by complex conjugation, 
so that $m_2^*=m_1\equiv m_R+im_I$.  
Then Eqs.~(7b) give 
$$|n_1|=|m_0-m_R+\sqrt{3}m_I|~,~~~
|n_2|=|m_0-m_R-\sqrt{3}m_I|~,~~~
|n_3|=|m_0+2m_R|~~~,\eqno(8a)$$
and we can satisfy $M_1/M_3=|n_1|/|n_3|$ and $M_2/M_3=|n_2|/|n_3|$ by taking 
$$m_R/m_0=1+{3 \over 2} {M_1+M_2\over M_3 -M_1-M_2}~,~~~
m_I/m_0={\sqrt{3}\over 2} {M_2-M_1 \over M_3-M_1-M_2}~~~.\eqno(8b)$$
Since this specialized form of $N$ can represent arbitrary mass ratios 
between the generations, so can the general form before specialization.  

Returning to the general form of N, let us determine the CKM matrix 
that results when we relax the assumption of partial symmetry, by allowing 
$P_{0,1,2}^f$ to have the general form of Eq.~(3b).  Thus, suppressing  
the superscript $f=u,d,e$ until needed, the 
mass matrix $M$ arising from Eq.~(4) now has the form
$$M=N+\epsilon~,~~~ \eqno(9a)$$
with $N$ given by Eq.~(5) and with $\epsilon$ a $3 \times 3$ complex matrix
with vanishing diagonal matrix elements (because $N$ has three independent 
parameters on its diagonal) and with arbitrary complex off-diagonal matrix 
elements. We diagonalize Eq.~(9a) in two steps.  First we apply the  
transformation of Eq.~(7a) that diagonalizes $N$, giving 
$$M^{\prime}=U_L (N+\epsilon) U_R^{\dagger}=
{\rm diag}(n_1,n_2,n_3) +\mu~,~~~\mu=U_L \epsilon U_R^{\dagger}~~~
.\eqno(9b)$$ 
Taking the inverse transformation $\epsilon=U_L^{\dagger} \mu U_R$, we 
easily find that the restrictions $\epsilon_{11}=\epsilon_{22}=\epsilon_{33}
=0$ translate into the restrictions 
$\mu_{11}=-\mu_{23}-\mu_{32}$, ~~$\mu_{22}=-\mu_{13}-\mu_{31}$,~~
$\mu_{33}=-\mu_{12}-\mu_{21}$ on the diagonal matrix elements of $\mu$.  
We now wish to find the additional 
left and right transformations $U_L^{\prime}$ and $U_R^{\prime \dagger}$  
that diagonalize $M^{\prime}$, treating 
$\mu$ as a perturbation.  Since calculation of the CKM matrix requires only 
knowledge of $U_L^{\prime}$, we first form the Hermitian matrix 
$M^{\prime}M^{\prime \dagger}$ and construct its diagonalizing unitary 
transformation.  From Eq.~(9b) we find 
$$M^{\prime}M^{\prime \dagger}={\rm diag}(|n_1|^2,|n_2|^2,|n_3|^2)+
\Delta~~~,\eqno(10a)$$
with $\Delta$ the Hermitian matrix with upper diagonal matrix elements 
given by  
$$\eqalign{
\Delta_{11}=&-2 {\rm Re}[n_1(\mu_{23}^*+\mu_{32}^*)]~,~~~
\Delta_{12}=n_1\mu_{21}^*+n_2^*\mu_{12}~,~~~
\Delta_{13}=n_1\mu_{31}^*+n_3^*\mu_{13}~~~,\cr
\Delta_{22}=&-2{\rm Re}[n_2(\mu_{13}^*+\mu_{31}^*)]~,~~~
\Delta_{23}=n_2\mu_{32}^*+n_3^*\mu_{23}~~~,\cr
\Delta_{33}=&-2{\rm Re}[n_3(\mu_{12}^*+\mu_{21}^*)]~~~.\cr
}\eqno(10b)$$
The problem of diagonalizing Eqs.(10a,~b) is textbook time-independent 
perturbation theory, and the solution is the matrix  
$$U_L^{\prime}=
\pmatrix{1 & {\Delta_{12}\over |n_1|^2-|n_2|^2}&
{\Delta_{13}\over|n_1|^2-|n_3|^2}\cr
{\Delta_{21}\over |n_2|^2-|n_1|^2} &1&{\Delta_{23}\over|n_2|^2-|n_3|^2}\cr
{\Delta_{31}\over|n_3|^2-|n_1|^2}&{\Delta_{32}\over|n_3|^2-|n_2|^2}&1\cr}~~~.
\eqno(11)$$
Restoring the superscripts $u,d$, the CKM mixing matrix is given by 
$U_{CKM}=U_L^{u \prime \dagger} U_L^{d \prime}$, which on using Eq.~(11) 
gives 
$$\eqalign{
U_{CKM}=&\pmatrix{1 & U_{12} & U_{13}\cr
                   -U_{12}^* & 1 & U_{23} \cr
                   -U_{13}^* & -U_{23}^* & 1\cr}~~~, \cr
U_{12}=&{\Delta^d_{12} \over |n_1^d|^2-|n_2^d|^2}
-{\Delta^u_{12} \over |n_1^u|^2-|n_2^u|^2}  ~~~,\cr
U_{13}=&{\Delta^d_{13} \over |n_1^d|^2-|n_3^d|^2}
-{\Delta^u_{13} \over |n_1^u|^2-|n_3^u|^2}  ~~~,\cr
U_{23}=&{\Delta^d_{23} \over |n_2^d|^2-|n_3^d|^2}
-{\Delta^u_{23} \over |n_2^u|^2-|n_3^u|^2}  ~~~.\cr
 }\eqno(12a) $$
To first order in perturbation theory, the quark masses are given by 
$$\eqalign{
M_u^2=|n_1^u|^2+\Delta_{11}^u~,~~~M_c^2=|n_2^u|^2+\Delta_{22}^u
~,~~~M_t^2=|n_3^u|^2+\Delta_{33}^u~~~,\cr
M_d^2=|n_1^d|^2+\Delta_{11}^d~,~~~M_s^2=|n_2^d|^2+\Delta_{22}^d
~,~~~M_b^2=|n_3^d|^2+\Delta_{33}^d~~~.\cr
}\eqno(12b)$$
Equations (12a~,b) are our final result for the CKM matrix, and the fermion 
masses, to first order in perturbation theory in the asymmetric 
perturbation $\mu$.

To examine the qualitative form of Eq.~(12a), let us make the approximation 
of neglecting the small quantities $M_d/M_s,M_s/M_c,M_s/M_b,M_b/M_t$, etc., 
and of neglecting the first order corrections of Eq.~(12b) to the 
masses $M_s$ and $M_b$, 
so that these are equal to $|n_2^d|$ and $|n_3^d|$ respectively.  Then 
combining the above formulas, and rephasing to put the CKM matrix in the 
standard form (to first order in the mixing angles) 
$$U_{CKM}=\pmatrix{1&s_{12}&s_{13}e^{-i\delta_{13}} \cr
                    -s_{12}&1&s_{23} \cr
                    -s_{13}e^{i\delta_{13}}& -s_{23}& 1 \cr}~~~,\eqno(13)$$
we find for the sines of the mixing angles
$$s_{12}={|\mu_{12}^d| \over M_s}~,~~~s_{13}={|\mu_{13}^d| \over M_b}
~,~~~s_{23}={|\mu_{23}^d| \over M_b }~~~,\eqno(14a)$$
and for the CP-violating phase
$$e^{-i\delta_{13}}=-{n_2^d \over |n_2^d|} {\mu_{12}^{d*} \over |\mu_{12}^d|}
{\mu_{13}^d \over |\mu_{13}^d|} {\mu_{23}^{d*}\over |\mu_{23}^d|}~~~.
\eqno(14b)$$
If the magnitudes of the off-diagonal matrix elements of $\mu$ are assumed 
roughly equal, then Eq.~(14a) tells us that the third family mixing angles 
are suppressed relative to the Cabibbo angle by a factor $M_s/M_b \sim 
1/20$, which to within factors of order unity (that can be accounted for 
by differences in the $\mu$'s) is in accord with experiment.  From the  
estimate of Eqs.~(8a~,b), we have $m_0^d \sim m_1^d \sim m_2^d\sim M_b/3$, 
with the 
fractional variation among them of order $M_s/M_b$.  Since the observed 
value of $s_{12}$ is $\sim 1/5$, we have $|\mu_{12}^d|/M_b\sim 1/100$. 
Thus the matrix elements of the initial 
partially symmetric approximation to the 
mass matrix are uniform to about five percent, and 
the additional fractional variation of the mass 
matrix elements supplied by the perturbation $\epsilon$ in Eq.~(9a), that 
breaks the partial symmetry, is of order three percent.   As a consequence,   
flavor changing neutral current effects arising from single Higgs exchange 
are suppressed by a factor of order $.03^2 \sim  10^{-3}$ relative to 
the generic estimate of Glashow and Weinberg [6], and one finds that for 
Higgs masses greater than roughly $300$ GeV, the Higgs 
exchange contribution 
to the $K_L-K_S$ mass difference is acceptably small.  

In a paper in preparation, we will give the full structure of the $Z_6$ 
invariant extension of the standard model referred to above.  Additionally,  
motivated by the model of Ref. [3], we will give a $Z_{12}$ 
invariant extension 
of the standard model that separates, under an $S(2)$ 
symmetry restriction, into two copies of the $Z_6$ model that interact 
only through the Higgs sector, and that have Yukawa couplings that are 
orthogonal linear combinations of the original $Z_{12}$ Yukawa couplings.  
Hence when the $Z_{12}$ model is close to the ``democratic'' limit, one 
of the $Z_6$ copies is also close, and will behave as discussed here; 
the second copy can have a very different (and possibly much lighter) 
mass spectrum.  In a second 
paper in preparation, we will investigate the possibilities for  
realizing a $Z_6$ or $Z_{12}$ discrete chiral symmetry structure 
in preonic models, again following the directions set out in Ref. [3].

\bigskip
\bigskip
\centerline{\bf Acknowledgments}
This work was supported in part by the Department of Energy under
Grant \#DE--FG02--90ER40542.  I wish to thank Henry Frisch, Harald Fritzsch, 
Chris Kolda, 
Jon Rosner, and Sam Treiman for stimulating conversations.  
\vfill\eject
\centerline{\bf References}
\bigskip
\noindent
\item{[1]} H. Harari, H. Haut, and J. Weyers, Phys. Lett. {\bf B78}, 459 
(1978); Y. Chikashige, G. Gelmini, R. P. Peccei, and M. 
Roncadelli, Phys. Lett.
{\bf B94}, 499 (1980); H. Fritzsch, in Proc. of Europhys. Conf. on Flavor 
Mixing in Weak Interactions, Erice, Italy (1984); C. Jarlskog, in Proc. of 
Int. Symp. on Production and Decay of Heavy Flavors, Heidelberg, Germany 
(1986); P. Kaus and S. Meshkov, Mod. Phys. Lett. {\bf A3}, 1251 (1988); 
Y. Koide, Phys. Rev. {\bf D39}, 1391 (1989); M. Tanimoto, Phys. Rev. 
{\bf D41}, 1586 (1990); G. C. Branco, J. I. Silva-Marcos, 
and M. N. Rebelo, Phys. Lett. 
{\bf B237}, 446 (1990); H. Fritzsch and J. Plankl, Phys. Lett. 
{\bf B237}, 451 (1990); H. Fritzsch, Phys. Lett. {\bf B289}, 92 (1992); 
H. Fritzsch and D. Holtmannsp\"otter, Phys. Lett. {\bf B338}, 290 (1994);
H. Fritzsch and Z. Z. Xing, Phys. Lett. {\bf B353}, 114 (1995); K. Kang 
and S. K. Kang, Phys. Rev. {\bf D56}, 1511 (1997). 
\bigskip 
\noindent
\item{[2]}  H. Harari and N. Seiberg, Phys. Lett. {\bf B102}, 263 (1981).
\bigskip
\noindent
\item{[3]}  S. Adler, hep-th/9610190 (unpublished).   
\bigskip
\noindent
\item{[4]}  S. Weinberg, Phys. Lett. {\bf B102}, 401 (1981). 
\bigskip                                                   
\noindent
\item{[5]}  M. Marcus, {\it Basic Theorems in Matrix Theory}, National 
Bureau of Standards Applied Mathematics Series no. 57 (1964), Sec. 2.13; 
H. L. Hamburger and M. E. Grimshaw, {\it Linear transformations in 
n-dimensional vector space}, Cambridge University Press (1951), pp. 94-96.  
\bigskip
\noindent
\item{[6]}  S. L. Glashow and S. Weinberg, Phys. Rev. {\bf D15}, 1958 (1977). 
\bigskip
\noindent
\bigskip
\noindent
\bigskip
\noindent
\bigskip
\noindent
\bigskip
\noindent
\bigskip
\noindent
\bigskip
\noindent
\bigskip
\noindent
\bigskip
\noindent
\bigskip
\noindent
\bigskip
\noindent
\bigskip
\noindent
\bigskip
\noindent
\bigskip
\noindent
\bigskip
\noindent
\vfill
\eject
\bigskip
\bye

Return-Path: adler@sns.ias.edu 
Received: from thunder.sns.ias.edu (thunder [198.138.243.12])
	by blackhole.sns.ias.edu (8.8.5/8.8.5) with ESMTP id OAA18030
	for <val@sns.ias.edu>; Mon, 24 Nov 1997 14:52:16 -0500 (EST)
Received: from thunder (adler@localhost)
	by thunder.sns.ias.edu (8.8.5/8.8.5) with ESMTP id OAA00531
	for <val>; Mon, 24 Nov 1997 14:52:13 -0500 (EST)
Message-Id: <199711241952.OAA00531@thunder.sns.ias.edu>
X-Authentication-Warning: thunder.sns.ias.edu: adler owned process doing -bs
X-Mailer: exmh version 1.6.7 5/3/96
To: val@sns.ias.edu
Mime-Version: 1.0
Date: Mon, 24 Nov 1997 14:52:13 -0500
From: Stephen Adler <adler@sns.ias.edu>
Content-Type: text/plain; charset=us-ascii
Content-Length: 32154

Please substitute this for the bulletin board version.  I corrected the 
problems you found, and have made some additions as well. 

Steve
_________________________________________-
\catcode`\@=11					% To make protected \def's

%************************************************************
%*
%*		Font set-up
%*
%************************************************************

%************** 5-point fonts *******************************

\font\fiverm=cmr5				% roman
\font\fivemi=cmmi5				% math italic
\font\fivesy=cmsy5				% math symbols
\font\fivebf=cmbx5				% bold face

\skewchar\fivemi='177
\skewchar\fivesy='60

%************** 6-point fonts *******************************

\font\sixrm=cmr6				% roman
\font\sixi=cmmi6				% math italic
\font\sixsy=cmsy6				% math symbols
\font\sixbf=cmbx6				% bold face

\skewchar\sixi='177
\skewchar\sixsy='60

%************** 7-point fonts *******************************

\font\sevenrm=cmr7				% roman
\font\seveni=cmmi7				% math italic
\font\sevensy=cmsy7				% math symbols
\font\sevenit=cmti7				% italic
\font\sevenbf=cmbx7				% bold face

\skewchar\seveni='177
\skewchar\sevensy='60

%************** 8-point fonts *******************************

\font\eightrm=cmr8				% roman
\font\eighti=cmmi8				% math italic
\font\eightsy=cmsy8				% math symbols
\font\eightit=cmti8				% italic
				% slanted
\font\eightbf=cmbx8				% bold face
				% typewriter
				% sans serif

\skewchar\eighti='177
\skewchar\eightsy='60

%************** 9-point fonts *******************************

\font\ninei=cmmi9
\font\ninesy=cmsy9

\skewchar\ninei='177
\skewchar\ninesy='60

%************** 10-point fonts ******************************

\font\tenrm=cmr10				% roman
\font\teni=cmmi10				% math italic
\font\tensy=cmsy10				% math symbols
\font\tenex=cmex10				% math extension
\font\tenit=cmti10				% italic
\font\tensl=cmsl10				% slanted
\font\tenbf=cmbx10				% bold face
\font\tentt=cmtt10				% typewriter
\font\tenss=cmss10				% sans serif
\font\tensc=cmcsc10				% small caps
\font\tenbi=cmmib10				% bold math

\skewchar\teni='177
\skewchar\tenbi='177
\skewchar\tensy='60

\def\tenpoint{\ifmmode\err@badsizechange\else
	\textfont0=\tenrm \scriptfont0=\sevenrm \scriptscriptfont0=\fiverm
	\textfont1=\teni  \scriptfont1=\seveni  \scriptscriptfont1=\fivemi
	\textfont2=\tensy \scriptfont2=\sevensy \scriptscriptfont2=\fivesy
	\textfont3=\tenex \scriptfont3=\tenex   \scriptscriptfont3=\tenex
	\textfont4=\tenit \scriptfont4=\sevenit \scriptscriptfont4=\sevenit
	\textfont5=\tensl
	\textfont6=\tenbf \scriptfont6=\sevenbf \scriptscriptfont6=\fivebf
	\textfont7=\tentt
	\textfont8=\tenbi \scriptfont8=\seveni  \scriptscriptfont8=\fivemi
	\def\rm{\tenrm\fam=0 }%
	\def\it{\tenit\fam=4 }%
	\def\sl{\tensl\fam=5 }%
	\def\bf{\tenbf\fam=6 }%
	\def\tt{\tentt\fam=7 }%
	\def\ss{\tenss}%
	\def\sc{\tensc}%
	\def\bmit{\fam=8 }%
	\rm\setparameters\setbaselines\fi}

%************** 12-point fonts ******************************

\font\twelverm=cmr12				% roman
\font\twelvei=cmmi12				% math italic
\font\twelvesy=cmsy10	scaled\magstep1		% math symbols
\font\twelveex=cmex10	scaled\magstep1		% math extension
\font\twelveit=cmti12				% italic
\font\twelvesl=cmsl12				% slanted
\font\twelvebf=cmbx12				% bold face
\font\twelvett=cmtt12				% typewriter
\font\twelvess=cmss12				% sans serif
\font\twelvesc=cmcsc10	scaled\magstep1		% small caps
\font\twelvebi=cmmib10	scaled\magstep1		% bold math

\skewchar\twelvei='177
\skewchar\twelvebi='177
\skewchar\twelvesy='60

\def\twelvepoint{\ifmmode\err@badsizechange\else
	\textfont0=\twelverm \scriptfont0=\eightrm \scriptscriptfont0=\sixrm
	\textfont1=\twelvei  \scriptfont1=\eighti  \scriptscriptfont1=\sixi
	\textfont2=\twelvesy \scriptfont2=\eightsy \scriptscriptfont2=\sixsy
	\textfont3=\twelveex \scriptfont3=\tenex   \scriptscriptfont3=\tenex
	\textfont4=\twelveit \scriptfont4=\eightit \scriptscriptfont4=\sevenit
	\textfont5=\twelvesl
	\textfont6=\twelvebf \scriptfont6=\eightbf \scriptscriptfont6=\sixbf
	\textfont7=\twelvett
	\textfont8=\twelvebi \scriptfont8=\eighti  \scriptscriptfont8=\sixi
	\def\rm{\twelverm\fam=0 }%
	\def\it{\twelveit\fam=4 }%
	\def\sl{\twelvesl\fam=5 }%
	\def\bf{\twelvebf\fam=6 }%
	\def\tt{\twelvett\fam=7 }%
	\def\ss{\twelvess}%
	\def\sc{\twelvesc}%
	\def\bmit{\fam=8 }%
	\rm\setparameters\setbaselines\fi}

%************** 14-point fonts ******************************

\font\fourteenrm=cmr12	scaled\magstep1		% roman
\font\fourteeni=cmmi12	scaled\magstep1		% math italic
\font\fourteensy=cmsy10	scaled\magstep2		% math symbols
\font\fourteenex=cmex10	scaled\magstep2		% math extension
\font\fourteenit=cmti12	scaled\magstep1		% italic
\font\fourteensl=cmsl12	scaled\magstep1		% slanted
\font\fourteenbf=cmbx12	scaled\magstep1		% bold face
\font\fourteentt=cmtt12	scaled\magstep1		% typewriter
\font\fourteenss=cmss12	scaled\magstep1		% sans serif
\font\fourteensc=cmcsc10 scaled\magstep2	% small caps
\font\fourteenbi=cmmib10 scaled\magstep2	% bold math

\skewchar\fourteeni='177
\skewchar\fourteenbi='177
\skewchar\fourteensy='60

\def\fourteenpoint{\ifmmode\err@badsizechange\else
	\textfont0=\fourteenrm \scriptfont0=\tenrm \scriptscriptfont0=\sevenrm
	\textfont1=\fourteeni  \scriptfont1=\teni  \scriptscriptfont1=\seveni
	\textfont2=\fourteensy \scriptfont2=\tensy \scriptscriptfont2=\sevensy
	\textfont3=\fourteenex \scriptfont3=\tenex \scriptscriptfont3=\tenex
	\textfont4=\fourteenit \scriptfont4=\tenit \scriptscriptfont4=\sevenit
	\textfont5=\fourteensl
	\textfont6=\fourteenbf \scriptfont6=\tenbf \scriptscriptfont6=\sevenbf
	\textfont7=\fourteentt
	\textfont8=\fourteenbi \scriptfont8=\tenbi \scriptscriptfont8=\seveni
	\def\rm{\fourteenrm\fam=0 }%
	\def\it{\fourteenit\fam=4 }%
	\def\sl{\fourteensl\fam=5 }%
	\def\bf{\fourteenbf\fam=6 }%
	\def\tt{\fourteentt\fam=7}%
	\def\ss{\fourteenss}%
	\def\sc{\fourteensc}%
	\def\bmit{\fam=8 }%
	\rm\setparameters\setbaselines\fi}

%************** Miscellaneous big fonts *********************

\font\seventeenrm=cmr10 scaled\magstep3		% roman
		% bold face

%************************************************************
%*
%*		Parameter initialization
%*
%************************************************************

\newdimen\rp@
\newcount\@basestretchnum
\newskip\@baseskip
\newskip\headskip
\newskip\footskip

% Routine to set page parameters

\def\setparameters{\rp@=.1em
	\headskip=24\rp@
	\footskip=\headskip
	\delimitershortfall=5\rp@
	\nulldelimiterspace=1.2\rp@
	\scriptspace=0.5\rp@
	\abovedisplayskip=10\rp@ plus3\rp@ minus5\rp@
	\belowdisplayskip=10\rp@ plus3\rp@ minus5\rp@
	\abovedisplayshortskip=5\rp@ plus2\rp@ minus4\rp@
	\belowdisplayshortskip=10\rp@ plus3\rp@ minus5\rp@
	\normallineskip=\rp@
	\lineskip=\normallineskip
	\normallineskiplimit=0pt
	\lineskiplimit=\normallineskiplimit
	\jot=3\rp@
	\setbox0=\hbox{\the\textfont3 B}\p@renwd=\wd0
	\skip\footins=12\rp@ plus3\rp@ minus3\rp@
	\skip\topins=0pt plus0pt minus0pt}

% Special routine to scale \baselineskip

\def\setbaselines{\maxdepth=4\rp@\baselinestretch=\@basestretchnum}

% The \baselinestretch command

\def\baselinestretch{\afterassignment\@basestretch\@basestretchnum}
\def\@basestretch{%
	\@baseskip=12\rp@ \divide\@baseskip by1000
	\normalbaselineskip=\@basestretchnum\@baseskip
	\baselineskip=\normalbaselineskip
	\bigskipamount=\the\baselineskip
		plus.25\baselineskip minus.25\baselineskip
	\medskipamount=.5\baselineskip
		plus.125\baselineskip minus.125\baselineskip
	\smallskipamount=.25\baselineskip
		plus.0625\baselineskip minus.0625\baselineskip
	\setbox\strutbox=\hbox{\vrule height.708\baselineskip
		depth.292\baselineskip width0pt }}

%************************************************************
%*
%*		Modifications to PLAIN.TEX
%*
%************************************************************

% Modifications to PLAIN routines to handle scaling of page parameters

\def\makeheadline{\vbox to0pt{\baselinestretch=1000
	\vskip-\headskip \vskip1.5pt
	\line{\vbox to\ht\strutbox{}\the\headline}\vss}\nointerlineskip}

\def\makefootline{\baselineskip=\footskip\line{\the\footline}}

\def\big#1{{\hbox{$\left#1\vbox to8.5\rp@ {}\right.\n@space$}}}
\def\Big#1{{\hbox{$\left#1\vbox to11.5\rp@ {}\right.\n@space$}}}
\def\bigg#1{{\hbox{$\left#1\vbox to14.5\rp@ {}\right.\n@space$}}}
\def\Bigg#1{{\hbox{$\left#1\vbox to17.5\rp@ {}\right.\n@space$}}}

% Modifications to PLAIN to handle bold math

\mathchardef\alpha="710B
\mathchardef\beta="710C
\mathchardef\gamma="710D
\mathchardef\delta="710E
\mathchardef\epsilon="710F
\mathchardef\zeta="7110
\mathchardef\eta="7111
\mathchardef\theta="7112
\mathchardef\iota="7113
\mathchardef\kappa="7114
\mathchardef\lambda="7115
\mathchardef\mu="7116
\mathchardef\nu="7117
\mathchardef\xi="7118
\mathchardef\pi="7119
\mathchardef\rho="711A
\mathchardef\sigma="711B
\mathchardef\tau="711C
\mathchardef\upsilon="711D
\mathchardef\phi="711E
\mathchardef\chi="711F
\mathchardef\psi="7120
\mathchardef\omega="7121
\mathchardef\varepsilon="7122
\mathchardef\vartheta="7123
\mathchardef\varpi="7124
\mathchardef\varrho="7125
\mathchardef\varsigma="7126
\mathchardef\varphi="7127
\mathchardef\imath="717B
\mathchardef\jmath="717C
\mathchardef\ell="7160
\mathchardef\wp="717D
\mathchardef\partial="7140
\mathchardef\flat="715B
\mathchardef\natural="715C
\mathchardef\sharp="715D

%************************************************************
%*
%*		Initialization
%*
%************************************************************

\def\err@badsizechange{%
	\immediate\write16{--> Size change not allowed in math mode, ignored}}

\baselinestretch=1000
\tenpoint

\catcode`\@=12					% Restore @ sign
% Routine to guarantee that this file is input only once
\catcode`\@=11
\expandafter\ifx\csname @iasmacros\endcsname\relax
	\global\let\@iasmacros=\par
\else	\immediate\write16{}
	\immediate\write16{Warning:}
	\immediate\write16{You have tried to input iasmacros more than once.}
	\immediate\write16{}
	\endinput
\fi
\catcode`\@=12

% Set up font size commands and \baselinestretch command
%\input iasfonts

% Some alternative font names
\def\rmb{\seventeenrm}

% Simple spacing commands
\def\singlespace{\baselineskip=\normalbaselineskip}
\def\halfspace{\baselineskip=1.5\normalbaselineskip}
\def\doublespace{\baselineskip=2\normalbaselineskip}

% Macros for references and abstracts

\def\AB{\bigskip\parindent=40pt
        \centerline{\bf ABSTRACT}\medskip\halfspace\narrower}
\def\AE{\bigskip\nonarrower\doublespace}
\def\nonarrower{\advance\leftskip by-\parindent
	\advance\rightskip by-\parindent}

% Useful commands

\def\boxit#1{\vbox{\hrule\hbox{\vrule\kern3pt
	\vbox{\kern3pt#1\kern3pt}\kern3pt\vrule}\hrule}}

% Special symbols
\def\hence{\leavevmode\hbox{\bf .\raise5.5pt\hbox{.}.} }

\def\dalemb#1#2{{\vbox{\hrule height.#2pt
	\hbox{\vrule width.#2pt height#1pt \kern#1pt \vrule width.#2pt}
	\hrule height.#2pt}}}
\def\gtorder{\mathrel{\raise.3ex\hbox{$>$}\mkern-14mu
             \lower0.6ex\hbox{$\sim$}}}
\def\ltorder{\mathrel{\raise.3ex\hbox{$<$}\mkern-14mu
             \lower0.6ex\hbox{$\sim$}}}

% For twoup output
\newdimen\fullhsize
\newbox\leftcolumn
\def\twoup{\hoffset=-.5in \voffset=-.25in
  \hsize=4.75in \fullhsize=10in \vsize=6.9in
  \def\fullline{\hbox to\fullhsize}
  \let\lr=L
  \output={\if L\lr
        \global\setbox\leftcolumn=\columnbox\global\let\lr=R \advancepageno
      \else \doubleformat \global\let\lr=L\fi
    \ifnum\outputpenalty>-20000 \else\dosupereject\fi}
  \def\doubleformat{\shipout\vbox{
    \fullline{\box\leftcolumn\hfil\columnbox}\advancepageno}}
  \def\columnbox{\leftline{\vbox{\makeheadline\pagebody\makefootline}}}
  \tolerance=1000 }

\twelvepoint
\doublespace
{\nopagenumbers{
\rightline{IASSNS-HEP-97/125}
\rightline{~~~November, 1997}
\bigskip\bigskip
\centerline{\rmb A Model for the Quark Mass and Flavor Mixing Matrices Based} 
\centerline{\rmb on Discrete Chiral Symmetry as the Origin of Families}
\medskip
\centerline{\it Stephen L. Adler}
\centerline{\bf Institute for Advanced Study}
\centerline{\bf Princeton, NJ 08540}
\medskip
\bigskip\bigskip
\leftline{\it Send correspondence to:}
\medskip
{\singlespace\leftline{Stephen L. Adler}
\leftline{Institute for Advanced Study}
\leftline{Olden Lane, Princeton, NJ 08540}
\leftline{Phone 609-734-8051; FAX 609-924-8399; email adler@ias.edu}}
\bigskip\bigskip
}}
\vfill\eject
\pageno=2
\AB
We construct a model for the quark mass and flavor mixing matrices, based on 
the hypothesis that the flavor weak eigenstates in the three families are 
distinguished by a spontaneously broken discrete $Z_6$ chiral symmetry.
In a leading partially symmetric approximation, the model accommodates 
the family 
mass spectra, with a CKM matrix that is exactly unity.  Adding small 
asymmetries in first order perturbation theory gives a CKM matrix with 
the correct qualitative structure, with $s_{23}/s_{12}$ and $s_{13}/s_{12}$ 
naturally of order $M_s/M_b$.  
\AE
\bigskip\bigskip
\vfill\eject
\pageno=3
It has long been recognized that the hierarchical structures of the 
family mass spectra, with their large third family masses, and of the CKM 
mixing matrix, with its suppressed third family mixings, may have a common
dynamical origin.  In particular, several authors [1] have stressed 
that the 
observed pattern seems to be close to the ``rank-one'' limit, in which the 
mass matrices have the ``democratic'' form of a matrix with all matrix 
elements equal to unity, which has one eigenvalue 1 and two eigenvalues 0;  
when both up and down quark mass matrices have this form, 
they are diagonalized 
by the same unitary transformation and the CKM matrix is unity.  However, 
because the underlying dynamical basis for this choice has not been 
apparent, it has not been possible to systematically extend the rank-one   
model to one that incorporates, and relates, the observed mass and mixing 
hierarchies.  

We present in this letter a model for the quark mass and flavor mixing 
matrices, based on the underlying dynamical assumption that the three 
flavor weak eigenstates are distinguished by different eigenvalues of 
a discrete chiral $Z_6$ quantum number.  The idea that a discrete chiral 
quantum number may underlie family structure was introduced originally  
by Harari and Seiberg [2], and was developed recently by 
the author [3] in a
modified form that we follow here.  Also of relevance is the remark of  
Weinberg [4] that an unbroken discrete chiral quantum number suffices 
to enforce the masslessness of fermionic states.  Extending the general  
framework of this earlier work, we postulate that
all {\it complex} fields carry a discrete chiral family quantum number.  
Since the Higgs scalars in the standard model are complex, we introduce a 
triplet of Higgs doublets that carry $Z_6$ quantum numbers, and that 
are coupled to the fermions by Yukawa couplings 
so that the Lagrangian is $Z_6$ invariant.  
Spontaneous symmetry breaking, in which the neutral members of the three
Higgs doublets acquire vacuum expectations, then gives the 
fermion mass matrices that form the basis for our detailed analysis.  

In the simplest form of a theory with a discrete chiral $Z_6$ symmetry, 
the fundamental Lagrangian, as augmented by the instanton-induced potential, 
is invariant under the simultaneous transformations 
$$\xi_L \to \xi_L \exp(-2\pi i/6)~,~~~\xi_R \to \xi_R \exp(2\pi i/6)~~~
\eqno(1a)$$
of the unification scale fermion fields $\xi$.  
The fields in the low energy 
effective Lagrangian are in general nonlinear functionals of the fundamental 
fields.  Fermionic effective fields must be odd monomials in the fundamental 
fields, and so can come in three varieties $\psi_n$ with the discrete 
chiral transformation law
$$\psi_{nL} \to \psi_{nL} \exp(-(2n+1)2\pi i/6)~,~~~
\psi_{nR} \to \psi_{nR} \exp((2n+1)2\pi i/6)~,~~~
n=0,1,2~~~,\eqno(1b)$$
while complex bosonic effective fields must be even monomials in 
the fundamental fermion fields, and so can 
also come in three varieties $\phi_n$ 
with the discrete chiral transformation law 
$$\phi_n \to \phi_n \exp(2n 2\pi i/6)~,~~~n=0,1,2~~~.\eqno(1c)$$
Gauge fields are real fields, and since the phase in Eq.~(1c) never takes 
the value $-1$, they come in only one variety transforming with phase 
unity under discrete chiral transformations.   (Two 
varieties of gauge fields become possible when the discrete chiral group is 
$Z_{12}$ rather than $Z_6$, since $-1$ is then present in the set of even 
phases.)  Thus the minimal $Z_6$ invariant extension of the standard  
model consists of a triplicated set of fermions, and a triplicated set of 
Higgs doublets, obeying the transformation laws of Eqs.~(1b) and (1c) 
respectively, together with the usual gauge bosons,  with the Lagrangian 
constructed to be $Z_6$ invariant.  

We shall give the full Lagrangian 
structure of this theory elsewhere; 
for our present purpose it suffices to state 
that all of the usual gauge sector results are unchanged. Hence we only   
exhibit the most general Yukawa coupling of the Higgs fields to the fermions,
which grouping the the fermions $\psi_n$ into a 3-component column vector, 
takes the form 
$$\overline{Q}_L \Phi^d d_R + \overline{Q}_L \Phi^u u_R+
\overline{\psi}_L \Phi^e e_R~~~+ {\rm adjoint}.\eqno(2)$$
Here $Q_L$ and $\psi_L$ denote the usual left handed quark and lepton 
doublets, as extended into three component column vectors of doublets 
to incorporate the discrete chiral symmetry.  Similarly, $d_R$, $u_R$, 
and $e_R$ denote the usual quark and lepton right handed singlets, 
as extended 
into three component column vectors.  The $3 \times 3$ matrices $\Phi^{d,u,e}$ 

are given by 
$$\Phi^{d,e}=\sum_s P_s^{d,e} \phi_{n(s)}~,~~~
\Phi^{u}=\sum_s P_s^u \tilde \phi_{\tilde n(s)}~~~,\eqno(3a)$$
with $\phi$ and $\tilde \phi$ denoting respectively the Higgs doublet and 
the charge conjugate Higgs doublet, with $n(s)~,~ \tilde n(s)$ 
denoting the unique discrete 
chiral Higgs component, for each $s=1,2,3$, that cancels the fermionic phase 
selected by the combination matrices $P_s$. The three possibilities for $P_s$ 
are given by (with $f$ denoting $d,u,e$) 
$$P_0^{f}=\pmatrix{0       & 0       & \alpha_{13}^f    \cr
                   0       & \alpha_{22}^f   &0        \cr
                   \alpha_{31}^f &0  &0   \cr}~,~~~
P_1^{f}=\pmatrix{\alpha_{11}^f &0 &0 \cr
                 0 & 0& \alpha_{23}^f \cr
                 0 & \alpha_{32}^f & 0 \cr}~,~~~
P_2^{f}=\pmatrix{0 & \alpha_{12}^f & 0 \cr
                 \alpha_{21}^f & 0 & 0 \cr
                 0 & 0 & \alpha_{33}^f \cr}~~~,
\eqno(3b)$$
with the $\alpha$'s complex numbers.  On spontaneous symmetry breaking, 
the neutral components of each Higgs field $\phi_n$ obtains a vacuum 
expectation value $v_n$, and the Yukawa coupling of Eq.~(2) gives 
the mass term 
$$\sum_s[\overline{d}_L P_s^d v_{n(s)} d_R 
+ \overline{u}_L P_s^u v_{\tilde  n(s)} u_R
+\overline{e}_L P_s^e v_{n(s)} e_R~+ {\rm adjoint}].\eqno(4)$$
Equation (4) is the starting point for our analysis.  

We observe that 
there are two specializations of Eq.~(4) that are of interest.  The first 
is what we term the {\it partially symmetric} limit, in which the three Higgs 
expectations $v_n$ are not equal, but in which the $\alpha$'s appearing 
in each $P_s^f$ are equal to a common value $\alpha_s^f$, 
so that each Higgs couples universally to all three 
combinations of fermions that have the same overall phase under discrete 
chiral transformation.  When the partially symmetric limit is further 
specialized to the completely symmetric limit in which the products 
$\alpha_s^f v_{n(s)}$  (for $f=d,e$)  and $\alpha_s^f v_{\tilde n(s)}$ 
(for $f=u$) 
are independent of $s$, the mass matrices of Eq.~(4) for each charge 
sector become 
proportional to the ``democratic'' mass matrix with all matrix elements 
equal to unity.  

In the partially symmetric limit, the mass matrices are all proportional 
to a matrix of the general form 
$$N=\pmatrix{m_1 & m_2 & m_0\cr   
           m_2 & m_0 & m_1 \cr
           m_0 & m_1 & m_2\cr} ~~~.\eqno(5)$$
This matrix is a {\it circulant} [5] (the rows are related to one another 
by successive cyclic permutation), and thus has the property that it 
can be diagonalized by           
left and right unitary transformation matrices that are {\it independent} of 
the values of the complex numbers $m_0$, $m_1$, and $m_2$.  To show this, 
we introduce the unitary matrices $U_L$ and $U_R^{\dagger}$ given by 
$$U_L={1 \over \sqrt{3}}\pmatrix{1 & \omega^2 & \omega \cr
                                 1 & \omega & \omega^2 \cr
                                 1 & 1 & 1 \cr}~,~~~
U_R^{\dagger}={1 \over \sqrt{3}}\pmatrix{1 & 1 & 1 \cr
                               \omega^2 & \omega & 1 \cr
                               \omega & \omega^2 & 1 \cr}~~~,
                               \eqno(6a)$$
with $\omega$ the complex cube root of unity,                                
$$\eqalign{
\omega=&-{1\over 2}+{\sqrt{3}\over 2} i~,~~~
\omega^2=\omega^*= -{1\over 2}-{\sqrt{3}\over 2} i~~~~ \cr
\omega^3=&1~,~~~1+\omega+\omega^2=0~,~~~i(\omega^2-\omega)=\sqrt{3} ~~~.\cr
}\eqno(6b)$$
We then find that 
$$U_L N U_R^{\dagger}=\pmatrix{ n_1 & 0 & 0 \cr
                                0   & n_2 & 0 \cr
                                0 & 0 & n_3 \cr}~~~,\eqno(7a)$$
with the eigenvalues $n_{1,2,3}$ given by                                 
$$n_1=m_1+\omega^2 m_2+\omega m_0 ~,~~~
n_2= m_1+\omega m_2+\omega^2 m_0~,~~~
n_3=m_1+m_2+m_0~~~.\eqno(7b)$$
Because the transformation of Eq.~(6a) is independent of the values of 
the matrix elements of $N$, in the partially symmetric limit changing  
basis from weak eigenstates to mass eigenstates diagonalizes not only the 
mass term of Eq.~(4), but also the Yukawa coupling of Eq.~(2).  Hence  
in the partially symmetric limit there are no flavor changing neutral 
currents arising from tree level Higgs exchange.

To see that the magnitudes $|n_1|,|n_2|,|n_3|$ can represent arbitrary 
first, second, and third generation masses $M_1,M_2,M_3$, let us specialize 
to the case when $m_0$ is real and 
$m_2$ and $m_1$ are related by complex conjugation, 
so that $m_2^*=m_1\equiv m_R+im_I$.  
Then Eqs.~(7b) give 
$$|n_1|=|m_0-m_R+\sqrt{3}m_I|~,~~~
|n_2|=|m_0-m_R-\sqrt{3}m_I|~,~~~
|n_3|=|m_0+2m_R|~~~,\eqno(8a)$$
and we can satisfy $M_1/M_3=|n_1|/|n_3|$ and $M_2/M_3=|n_2|/|n_3|$ by taking 
$$m_R/m_0=1+{3 \over 2} {M_1+M_2\over M_3 -M_1-M_2}~,~~~
m_I/m_0={\sqrt{3}\over 2} {M_2-M_1 \over M_3-M_1-M_2}~~~.\eqno(8b)$$
Since this specialized form of $N$ can represent arbitrary mass ratios 
between the generations, so can the general form before specialization.  

Returning to the general form of N, let us determine the CKM matrix 
that results when we relax the assumption of partial symmetry, by allowing 
$P_{0,1,2}^f$ to have the general form of Eq.~(3b).  Thus, suppressing  
the superscript $f=u,d,e$ until needed, the 
mass matrix $M$ arising from Eq.~(4) now has the form
$$M=N+\epsilon~,~~~ \eqno(9a)$$
with $N$ given by Eq.~(5) and with $\epsilon$ a $3 \times 3$ complex matrix
with vanishing diagonal matrix elements (because $N$ has three independent 
parameters on its diagonal) and with arbitrary complex off-diagonal matrix 
elements. We diagonalize Eq.~(9a) in two steps.  First we apply the  
transformation of Eq.~(7a) that diagonalizes $N$, giving 
$$M^{\prime}=U_L (N+\epsilon) U_R^{\dagger}=
{\rm diag}(n_1,n_2,n_3) +\mu~,~~~\mu=U_L \epsilon U_R^{\dagger}~~~
.\eqno(9b)$$ 
Taking the inverse transformation $\epsilon=U_L^{\dagger} \mu U_R$, we 
easily find that the restrictions $\epsilon_{11}=\epsilon_{22}=\epsilon_{33}
=0$ translate into the restrictions 
$\mu_{11}=-\mu_{23}-\mu_{32}$, ~~$\mu_{22}=-\mu_{13}-\mu_{31}$,~~
$\mu_{33}=-\mu_{12}-\mu_{21}$ on the diagonal matrix elements of $\mu$.  
We now wish to find the additional 
left and right transformations $U_L^{\prime}$ and $U_R^{\prime \dagger}$  
that diagonalize $M^{\prime}$, treating 
$\mu$ as a perturbation.  Since calculation of the CKM matrix requires only 
knowledge of $U_L^{\prime}$, we first form the Hermitian matrix 
$M^{\prime}M^{\prime \dagger}$ and construct its diagonalizing unitary 
transformation.  From Eq.~(9b) we find 
$$M^{\prime}M^{\prime \dagger}={\rm diag}(|n_1|^2,|n_2|^2,|n_3|^2)+
\Delta~~~,\eqno(10a)$$
with $\Delta$ the Hermitian matrix with upper diagonal matrix elements 
given by  
$$\eqalign{
\Delta_{11}=&-2 {\rm Re}[n_1(\mu_{23}^*+\mu_{32}^*)]~,~~~
\Delta_{12}=n_1\mu_{21}^*+n_2^*\mu_{12}~,~~~
\Delta_{13}=n_1\mu_{31}^*+n_3^*\mu_{13}~~~,\cr
\Delta_{22}=&-2{\rm Re}[n_2(\mu_{13}^*+\mu_{31}^*)]~,~~~
\Delta_{23}=n_2\mu_{32}^*+n_3^*\mu_{23}~~~,\cr
\Delta_{33}=&-2{\rm Re}[n_3(\mu_{12}^*+\mu_{21}^*)]~~~.\cr
}\eqno(10b)$$
The problem of diagonalizing Eqs.(10a,~b) is textbook time-independent 
perturbation theory, and the solution is the matrix  
$$U_L^{\prime}=
\pmatrix{1 & {\Delta_{12}\over |n_1|^2-|n_2|^2}&
{\Delta_{13}\over|n_1|^2-|n_3|^2}\cr
{\Delta_{21}\over |n_2|^2-|n_1|^2} &1&{\Delta_{23}\over|n_2|^2-|n_3|^2}\cr
{\Delta_{31}\over|n_3|^2-|n_1|^2}&{\Delta_{32}\over|n_3|^2-|n_2|^2}&1\cr}~~~.
\eqno(11)$$
Restoring the superscripts $u,d$, the CKM mixing matrix is given by 
$U_{CKM}=U_L^{u \prime \dagger} U_L^{d \prime}$, which on using Eq.~(11) 
gives 
$$\eqalign{
U_{CKM}=&\pmatrix{1 & U_{12} & U_{13}\cr
                   -U_{12}^* & 1 & U_{23} \cr
                   -U_{13}^* & -U_{23}^* & 1\cr}~~~, \cr
U_{12}=&{\Delta^d_{12} \over |n_1^d|^2-|n_2^d|^2}
-{\Delta^u_{12} \over |n_1^u|^2-|n_2^u|^2}  ~~~,\cr
U_{13}=&{\Delta^d_{13} \over |n_1^d|^2-|n_3^d|^2}
-{\Delta^u_{13} \over |n_1^u|^2-|n_3^u|^2}  ~~~,\cr
U_{23}=&{\Delta^d_{23} \over |n_2^d|^2-|n_3^d|^2}
-{\Delta^u_{23} \over |n_2^u|^2-|n_3^u|^2}  ~~~.\cr
 }\eqno(12a) $$
To first order in perturbation theory, the quark masses are given by 
$$\eqalign{
M_u^2=|n_1^u|^2+\Delta_{11}^u~,~~~M_c^2=|n_2^u|^2+\Delta_{22}^u
~,~~~M_t^2=|n_3^u|^2+\Delta_{33}^u~~~,\cr
M_d^2=|n_1^d|^2+\Delta_{11}^d~,~~~M_s^2=|n_2^d|^2+\Delta_{22}^d
~,~~~M_b^2=|n_3^d|^2+\Delta_{33}^d~~~.\cr
}\eqno(12b)$$
Equations (12a~,b) are our final result for the CKM matrix, and the fermion 
masses, to first order in perturbation theory in the asymmetric 
perturbation $\mu$.

To examine the qualitative form of Eq.~(12a), let us make the approximation 
of neglecting the small quantities $M_d/M_s,M_s/M_c,M_s/M_b,M_b/M_t$, etc., 
and of neglecting the first order corrections of Eq.~(12b) to the 
masses $M_s$ and $M_b$, 
so that these are equal to $|n_2^d|$ and $|n_3^d|$ respectively.  Then 
combining the above formulas, and rephasing to put the CKM matrix in the 
standard form (to first order in the mixing angles) 
$$U_{CKM}=\pmatrix{1&s_{12}&s_{13}e^{-i\delta_{13}} \cr
                    -s_{12}&1&s_{23} \cr
                    -s_{13}e^{i\delta_{13}}& -s_{23}& 1 \cr}~~~,\eqno(13)$$
we find for the sines of the mixing angles
$$s_{12}={|\mu_{12}^d| \over M_s}~,~~~s_{13}={|\mu_{13}^d| \over M_b}
~,~~~s_{23}={|\mu_{23}^d| \over M_b }~~~,\eqno(14a)$$
and for the CP-violating phase
$$e^{-i\delta_{13}}=-{n_2^d \over |n_2^d|} {\mu_{12}^{d*} \over |\mu_{12}^d|}
{\mu_{13}^d \over |\mu_{13}^d|} {\mu_{23}^{d*}\over |\mu_{23}^d|}~~~.
\eqno(14b)$$
If the magnitudes of the off-diagonal matrix elements of $\mu$ are assumed 
roughly equal, then Eq.~(14a) tells us that the third family mixing angles 
are suppressed relative to the Cabibbo angle by a factor $M_s/M_b \sim 
1/20$, which to within factors of order unity (that can be accounted for 
by differences in the $\mu$'s) is in accord with experiment.  From the  
estimate of Eqs.~(8a~,b), we have $m_0^d \sim m_1^d \sim m_2^d\sim M_b/3$, 
with the 
fractional variation among them of order $M_s/M_b$.  Since the observed 
value of $s_{12}$ is $\sim 1/5$, we have $|\mu_{12}^d|/M_b\sim 1/100$. 
Thus the matrix elements of the initial 
partially symmetric approximation to the 
mass matrix are uniform to about five percent, and 
the additional fractional variation of the mass 
matrix elements supplied by the perturbation $\epsilon$ in Eq.~(9a), that 
breaks the partial symmetry, is of order three percent.   As a consequence,   
flavor changing neutral current effects arising from single Higgs exchange 
are suppressed by a factor of order $.03^2 \sim  10^{-3}$ relative to 
the generic estimate of Glashow and Weinberg [6], and one finds that for 
Higgs masses greater than roughly $300$ GeV, the Higgs 
exchange contribution 
to the $K_L-K_S$ mass difference is acceptably small.  

In a paper in preparation, we will give the full structure of the $Z_6$ 
invariant extension of the standard model referred to above.  Additionally,  
motivated by the model of Ref. [3], we will give a $Z_{12}$ 
invariant extension 
of the standard model that separates, under an $S(2)$ 
symmetry restriction, into two copies of the $Z_6$ model that interact 
only through the Higgs sector, and that have Yukawa couplings that are 
orthogonal linear combinations of the original $Z_{12}$ Yukawa couplings.  
Hence when the $Z_{12}$ model is close to the ``democratic'' limit, one 
of the $Z_6$ copies is also close, and will behave as discussed here; 
the second copy can have a very different (and possibly much lighter) 
mass spectrum.  In a second 
paper in preparation, we will investigate the possibilities for  
realizing a $Z_6$ or $Z_{12}$ discrete chiral symmetry structure 
in preonic models, again following the directions set out in Ref. [3].

\bigskip
\bigskip
\centerline{\bf Acknowledgments}
This work was supported in part by the Department of Energy under
Grant \#DE--FG02--90ER40542.  I wish to thank Henry Frisch, Harald Fritzsch, 
Chris Kolda, 
Jon Rosner, and Sam Treiman for stimulating conversations.  
\vfill\eject
\centerline{\bf References}
\bigskip
\noindent
\item{[1]} H. Harari, H. Haut, and J. Weyers, Phys. Lett. {\bf B78}, 459 
(1978); Y. Chikashige, G. Gelmini, R. P. Peccei, and M. 
Roncadelli, Phys. Lett.
{\bf B94}, 499 (1980); H. Fritzsch, in Proc. of Europhys. Conf. on Flavor 
Mixing in Weak Interactions, Erice, Italy (1984); C. Jarlskog, in Proc. of 
Int. Symp. on Production and Decay of Heavy Flavors, Heidelberg, Germany 
(1986); P. Kaus and S. Meshkov, Mod. Phys. Lett. {\bf A3}, 1251 (1988); 
Y. Koide, Phys. Rev. {\bf D39}, 1391 (1989); M. Tanimoto, Phys. Rev. 
{\bf D41}, 1586 (1990); G. C. Branco, J. I. Silva-Marcos, 
and M. N. Rebelo, Phys. Lett. 
{\bf B237}, 446 (1990); H. Fritzsch and J. Plankl, Phys. Lett. 
{\bf B237}, 451 (1990); H. Fritzsch, Phys. Lett. {\bf B289}, 92 (1992); 
H. Fritzsch and D. Holtmannsp\"otter, Phys. Lett. {\bf B338}, 290 (1994);
H. Fritzsch and Z. Z. Xing, Phys. Lett. {\bf B353}, 114 (1995); K. Kang 
and S. K. Kang, Phys. Rev. {\bf D56}, 1511 (1997). 
\bigskip 
\noindent
\item{[2]}  H. Harari and N. Seiberg, Phys. Lett. {\bf B102}, 263 (1981).
\bigskip
\noindent
\item{[3]}  S. Adler, hep-th/9610190 (unpublished).   
\bigskip
\noindent
\item{[4]}  S. Weinberg, Phys. Lett. {\bf B102}, 401 (1981). 
\bigskip                                                   
\noindent
\item{[5]}  M. Marcus, {\it Basic Theorems in Matrix Theory}, National 
Bureau of Standards Applied Mathematics Series no. 57 (1964), Sec. 2.13; 
H. L. Hamburger and M. E. Grimshaw, {\it Linear transformations in 
n-dimensional vector space}, Cambridge University Press (1951), pp. 94-96.  
\bigskip
\noindent
\item{[6]}  S. L. Glashow and S. Weinberg, Phys. Rev. {\bf D15}, 1958 (1977). 
\bigskip
\noindent
\bigskip
\noindent
\bigskip
\noindent
\bigskip
\noindent
\bigskip
\noindent
\bigskip
\noindent
\bigskip
\noindent
\bigskip
\noindent
\bigskip
\noindent
\bigskip
\noindent
\bigskip
\noindent
\bigskip
\noindent
\bigskip
\noindent
\bigskip
\noindent
\bigskip
\noindent
\vfill
\eject
\bigskip
\bye

Return-Path: adler@sns.ias.edu 
Received: from thunder.sns.ias.edu (thunder [198.138.243.12])
	by blackhole.sns.ias.edu (8.8.5/8.8.5) with ESMTP id OAA18030
	for <val@sns.ias.edu>; Mon, 24 Nov 1997 14:52:16 -0500 (EST)
Received: from thunder (adler@localhost)
	by thunder.sns.ias.edu (8.8.5/8.8.5) with ESMTP id OAA00531
	for <val>; Mon, 24 Nov 1997 14:52:13 -0500 (EST)
Message-Id: <199711241952.OAA00531@thunder.sns.ias.edu>
X-Authentication-Warning: thunder.sns.ias.edu: adler owned process doing -bs
X-Mailer: exmh version 1.6.7 5/3/96
To: val@sns.ias.edu
Mime-Version: 1.0
Date: Mon, 24 Nov 1997 14:52:13 -0500
From: Stephen Adler <adler@sns.ias.edu>
Content-Type: text/plain; charset=us-ascii
Content-Length: 32154

Please substitute this for the bulletin board version.  I corrected the 
problems you found, and have made some additions as well. 

Steve
_________________________________________-
\catcode`\@=11					% To make protected \def's

%************************************************************
%*
%*		Font set-up
%*
%************************************************************

%************** 5-point fonts *******************************

\font\fiverm=cmr5				% roman
\font\fivemi=cmmi5				% math italic
\font\fivesy=cmsy5				% math symbols
\font\fivebf=cmbx5				% bold face

\skewchar\fivemi='177
\skewchar\fivesy='60

%************** 6-point fonts *******************************

\font\sixrm=cmr6				% roman
\font\sixi=cmmi6				% math italic
\font\sixsy=cmsy6				% math symbols
\font\sixbf=cmbx6				% bold face

\skewchar\sixi='177
\skewchar\sixsy='60

%************** 7-point fonts *******************************

\font\sevenrm=cmr7				% roman
\font\seveni=cmmi7				% math italic
\font\sevensy=cmsy7				% math symbols
\font\sevenit=cmti7				% italic
\font\sevenbf=cmbx7				% bold face

\skewchar\seveni='177
\skewchar\sevensy='60

%************** 8-point fonts *******************************

\font\eightrm=cmr8				% roman
\font\eighti=cmmi8				% math italic
\font\eightsy=cmsy8				% math symbols
\font\eightit=cmti8				% italic
				% slanted
\font\eightbf=cmbx8				% bold face
				% typewriter
				% sans serif

\skewchar\eighti='177
\skewchar\eightsy='60

%************** 9-point fonts *******************************

\font\ninei=cmmi9
\font\ninesy=cmsy9

\skewchar\ninei='177
\skewchar\ninesy='60

%************** 10-point fonts ******************************

\font\tenrm=cmr10				% roman
\font\teni=cmmi10				% math italic
\font\tensy=cmsy10				% math symbols
\font\tenex=cmex10				% math extension
\font\tenit=cmti10				% italic
\font\tensl=cmsl10				% slanted
\font\tenbf=cmbx10				% bold face
\font\tentt=cmtt10				% typewriter
\font\tenss=cmss10				% sans serif
\font\tensc=cmcsc10				% small caps
\font\tenbi=cmmib10				% bold math

\skewchar\teni='177
\skewchar\tenbi='177
\skewchar\tensy='60

\def\tenpoint{\ifmmode\err@badsizechange\else
	\textfont0=\tenrm \scriptfont0=\sevenrm \scriptscriptfont0=\fiverm
	\textfont1=\teni  \scriptfont1=\seveni  \scriptscriptfont1=\fivemi
	\textfont2=\tensy \scriptfont2=\sevensy \scriptscriptfont2=\fivesy
	\textfont3=\tenex \scriptfont3=\tenex   \scriptscriptfont3=\tenex
	\textfont4=\tenit \scriptfont4=\sevenit \scriptscriptfont4=\sevenit
	\textfont5=\tensl
	\textfont6=\tenbf \scriptfont6=\sevenbf \scriptscriptfont6=\fivebf
	\textfont7=\tentt
	\textfont8=\tenbi \scriptfont8=\seveni  \scriptscriptfont8=\fivemi
	\def\rm{\tenrm\fam=0 }%
	\def\it{\tenit\fam=4 }%
	\def\sl{\tensl\fam=5 }%
	\def\bf{\tenbf\fam=6 }%
	\def\tt{\tentt\fam=7 }%
	\def\ss{\tenss}%
	\def\sc{\tensc}%
	\def\bmit{\fam=8 }%
	\rm\setparameters\setbaselines\fi}

%************** 12-point fonts ******************************

\font\twelverm=cmr12				% roman
\font\twelvei=cmmi12				% math italic
\font\twelvesy=cmsy10	scaled\magstep1		% math symbols
\font\twelveex=cmex10	scaled\magstep1		% math extension
\font\twelveit=cmti12				% italic
\font\twelvesl=cmsl12				% slanted
\font\twelvebf=cmbx12				% bold face
\font\twelvett=cmtt12				% typewriter
\font\twelvess=cmss12				% sans serif
\font\twelvesc=cmcsc10	scaled\magstep1		% small caps
\font\twelvebi=cmmib10	scaled\magstep1		% bold math

\skewchar\twelvei='177
\skewchar\twelvebi='177
\skewchar\twelvesy='60

\def\twelvepoint{\ifmmode\err@badsizechange\else
	\textfont0=\twelverm \scriptfont0=\eightrm \scriptscriptfont0=\sixrm
	\textfont1=\twelvei  \scriptfont1=\eighti  \scriptscriptfont1=\sixi
	\textfont2=\twelvesy \scriptfont2=\eightsy \scriptscriptfont2=\sixsy
	\textfont3=\twelveex \scriptfont3=\tenex   \scriptscriptfont3=\tenex
	\textfont4=\twelveit \scriptfont4=\eightit \scriptscriptfont4=\sevenit
	\textfont5=\twelvesl
	\textfont6=\twelvebf \scriptfont6=\eightbf \scriptscriptfont6=\sixbf
	\textfont7=\twelvett
	\textfont8=\twelvebi \scriptfont8=\eighti  \scriptscriptfont8=\sixi
	\def\rm{\twelverm\fam=0 }%
	\def\it{\twelveit\fam=4 }%
	\def\sl{\twelvesl\fam=5 }%
	\def\bf{\twelvebf\fam=6 }%
	\def\tt{\twelvett\fam=7 }%
	\def\ss{\twelvess}%
	\def\sc{\twelvesc}%
	\def\bmit{\fam=8 }%
	\rm\setparameters\setbaselines\fi}

%************** 14-point fonts ******************************

\font\fourteenrm=cmr12	scaled\magstep1		% roman
\font\fourteeni=cmmi12	scaled\magstep1		% math italic
\font\fourteensy=cmsy10	scaled\magstep2		% math symbols
\font\fourteenex=cmex10	scaled\magstep2		% math extension
\font\fourteenit=cmti12	scaled\magstep1		% italic
\font\fourteensl=cmsl12	scaled\magstep1		% slanted
\font\fourteenbf=cmbx12	scaled\magstep1		% bold face
\font\fourteentt=cmtt12	scaled\magstep1		% typewriter
\font\fourteenss=cmss12	scaled\magstep1		% sans serif
\font\fourteensc=cmcsc10 scaled\magstep2	% small caps
\font\fourteenbi=cmmib10 scaled\magstep2	% bold math

\skewchar\fourteeni='177
\skewchar\fourteenbi='177
\skewchar\fourteensy='60

\def\fourteenpoint{\ifmmode\err@badsizechange\else
	\textfont0=\fourteenrm \scriptfont0=\tenrm \scriptscriptfont0=\sevenrm
	\textfont1=\fourteeni  \scriptfont1=\teni  \scriptscriptfont1=\seveni
	\textfont2=\fourteensy \scriptfont2=\tensy \scriptscriptfont2=\sevensy
	\textfont3=\fourteenex \scriptfont3=\tenex \scriptscriptfont3=\tenex
	\textfont4=\fourteenit \scriptfont4=\tenit \scriptscriptfont4=\sevenit
	\textfont5=\fourteensl
	\textfont6=\fourteenbf \scriptfont6=\tenbf \scriptscriptfont6=\sevenbf
	\textfont7=\fourteentt
	\textfont8=\fourteenbi \scriptfont8=\tenbi \scriptscriptfont8=\seveni
	\def\rm{\fourteenrm\fam=0 }%
	\def\it{\fourteenit\fam=4 }%
	\def\sl{\fourteensl\fam=5 }%
	\def\bf{\fourteenbf\fam=6 }%
	\def\tt{\fourteentt\fam=7}%
	\def\ss{\fourteenss}%
	\def\sc{\fourteensc}%
	\def\bmit{\fam=8 }%
	\rm\setparameters\setbaselines\fi}

%************** Miscellaneous big fonts *********************

\font\seventeenrm=cmr10 scaled\magstep3		% roman
		% bold face

%************************************************************
%*
%*		Parameter initialization
%*
%************************************************************

\newdimen\rp@
\newcount\@basestretchnum
\newskip\@baseskip
\newskip\headskip
\newskip\footskip

% Routine to set page parameters

\def\setparameters{\rp@=.1em
	\headskip=24\rp@
	\footskip=\headskip
	\delimitershortfall=5\rp@
	\nulldelimiterspace=1.2\rp@
	\scriptspace=0.5\rp@
	\abovedisplayskip=10\rp@ plus3\rp@ minus5\rp@
	\belowdisplayskip=10\rp@ plus3\rp@ minus5\rp@
	\abovedisplayshortskip=5\rp@ plus2\rp@ minus4\rp@
	\belowdisplayshortskip=10\rp@ plus3\rp@ minus5\rp@
	\normallineskip=\rp@
	\lineskip=\normallineskip
	\normallineskiplimit=0pt
	\lineskiplimit=\normallineskiplimit
	\jot=3\rp@
	\setbox0=\hbox{\the\textfont3 B}\p@renwd=\wd0
	\skip\footins=12\rp@ plus3\rp@ minus3\rp@
	\skip\topins=0pt plus0pt minus0pt}

% Special routine to scale \baselineskip

\def\setbaselines{\maxdepth=4\rp@\baselinestretch=\@basestretchnum}

% The \baselinestretch command

\def\baselinestretch{\afterassignment\@basestretch\@basestretchnum}
\def\@basestretch{%
	\@baseskip=12\rp@ \divide\@baseskip by1000
	\normalbaselineskip=\@basestretchnum\@baseskip
	\baselineskip=\normalbaselineskip
	\bigskipamount=\the\baselineskip
		plus.25\baselineskip minus.25\baselineskip
	\medskipamount=.5\baselineskip
		plus.125\baselineskip minus.125\baselineskip
	\smallskipamount=.25\baselineskip
		plus.0625\baselineskip minus.0625\baselineskip
	\setbox\strutbox=\hbox{\vrule height.708\baselineskip
		depth.292\baselineskip width0pt }}

%************************************************************
%*
%*		Modifications to PLAIN.TEX
%*
%************************************************************

% Modifications to PLAIN routines to handle scaling of page parameters

\def\makeheadline{\vbox to0pt{\baselinestretch=1000
	\vskip-\headskip \vskip1.5pt
	\line{\vbox to\ht\strutbox{}\the\headline}\vss}\nointerlineskip}

\def\makefootline{\baselineskip=\footskip\line{\the\footline}}

\def\big#1{{\hbox{$\left#1\vbox to8.5\rp@ {}\right.\n@space$}}}
\def\Big#1{{\hbox{$\left#1\vbox to11.5\rp@ {}\right.\n@space$}}}
\def\bigg#1{{\hbox{$\left#1\vbox to14.5\rp@ {}\right.\n@space$}}}
\def\Bigg#1{{\hbox{$\left#1\vbox to17.5\rp@ {}\right.\n@space$}}}

% Modifications to PLAIN to handle bold math

\mathchardef\alpha="710B
\mathchardef\beta="710C
\mathchardef\gamma="710D
\mathchardef\delta="710E
\mathchardef\epsilon="710F
\mathchardef\zeta="7110
\mathchardef\eta="7111
\mathchardef\theta="7112
\mathchardef\iota="7113
\mathchardef\kappa="7114
\mathchardef\lambda="7115
\mathchardef\mu="7116
\mathchardef\nu="7117
\mathchardef\xi="7118
\mathchardef\pi="7119
\mathchardef\rho="711A
\mathchardef\sigma="711B
\mathchardef\tau="711C
\mathchardef\upsilon="711D
\mathchardef\phi="711E
\mathchardef\chi="711F
\mathchardef\psi="7120
\mathchardef\omega="7121
\mathchardef\varepsilon="7122
\mathchardef\vartheta="7123
\mathchardef\varpi="7124
\mathchardef\varrho="7125
\mathchardef\varsigma="7126
\mathchardef\varphi="7127
\mathchardef\imath="717B
\mathchardef\jmath="717C
\mathchardef\ell="7160
\mathchardef\wp="717D
\mathchardef\partial="7140
\mathchardef\flat="715B
\mathchardef\natural="715C
\mathchardef\sharp="715D

%************************************************************
%*
%*		Initialization
%*
%************************************************************

\def\err@badsizechange{%
	\immediate\write16{--> Size change not allowed in math mode, ignored}}

\baselinestretch=1000
\tenpoint

\catcode`\@=12					% Restore @ sign
% Routine to guarantee that this file is input only once
\catcode`\@=11
\expandafter\ifx\csname @iasmacros\endcsname\relax
	\global\let\@iasmacros=\par
\else	\immediate\write16{}
	\immediate\write16{Warning:}
	\immediate\write16{You have tried to input iasmacros more than once.}
	\immediate\write16{}
	\endinput
\fi
\catcode`\@=12

% Set up font size commands and \baselinestretch command
%\input iasfonts

% Some alternative font names
\def\rmb{\seventeenrm}

% Simple spacing commands
\def\singlespace{\baselineskip=\normalbaselineskip}
\def\halfspace{\baselineskip=1.5\normalbaselineskip}
\def\doublespace{\baselineskip=2\normalbaselineskip}

% Macros for references and abstracts

\def\AB{\bigskip\parindent=40pt
        \centerline{\bf ABSTRACT}\medskip\halfspace\narrower}
\def\AE{\bigskip\nonarrower\doublespace}
\def\nonarrower{\advance\leftskip by-\parindent
	\advance\rightskip by-\parindent}

% Useful commands

\def\boxit#1{\vbox{\hrule\hbox{\vrule\kern3pt
	\vbox{\kern3pt#1\kern3pt}\kern3pt\vrule}\hrule}}

% Special symbols
\def\hence{\leavevmode\hbox{\bf .\raise5.5pt\hbox{.}.} }

\def\dalemb#1#2{{\vbox{\hrule height.#2pt
	\hbox{\vrule width.#2pt height#1pt \kern#1pt \vrule width.#2pt}
	\hrule height.#2pt}}}
\def\gtorder{\mathrel{\raise.3ex\hbox{$>$}\mkern-14mu
             \lower0.6ex\hbox{$\sim$}}}
\def\ltorder{\mathrel{\raise.3ex\hbox{$<$}\mkern-14mu
             \lower0.6ex\hbox{$\sim$}}}

% For twoup output
\newdimen\fullhsize
\newbox\leftcolumn
\def\twoup{\hoffset=-.5in \voffset=-.25in
  \hsize=4.75in \fullhsize=10in \vsize=6.9in
  \def\fullline{\hbox to\fullhsize}
  \let\lr=L
  \output={\if L\lr
        \global\setbox\leftcolumn=\columnbox\global\let\lr=R \advancepageno
      \else \doubleformat \global\let\lr=L\fi
    \ifnum\outputpenalty>-20000 \else\dosupereject\fi}
  \def\doubleformat{\shipout\vbox{
    \fullline{\box\leftcolumn\hfil\columnbox}\advancepageno}}
  \def\columnbox{\leftline{\vbox{\makeheadline\pagebody\makefootline}}}
  \tolerance=1000 }

\twelvepoint
\doublespace
{\nopagenumbers{
\rightline{IASSNS-HEP-97/125}
\rightline{~~~November, 1997}
\bigskip\bigskip
\centerline{\rmb A Model for the Quark Mass and Flavor Mixing Matrices Based} 
\centerline{\rmb on Discrete Chiral Symmetry as the Origin of Families}
\medskip
\centerline{\it Stephen L. Adler}
\centerline{\bf Institute for Advanced Study}
\centerline{\bf Princeton, NJ 08540}
\medskip
\bigskip\bigskip
\leftline{\it Send correspondence to:}
\medskip
{\singlespace\leftline{Stephen L. Adler}
\leftline{Institute for Advanced Study}
\leftline{Olden Lane, Princeton, NJ 08540}
\leftline{Phone 609-734-8051; FAX 609-924-8399; email adler@ias.edu}}
\bigskip\bigskip
}}
\vfill\eject
\pageno=2
\AB
We construct a model for the quark mass and flavor mixing matrices, based on 
the hypothesis that the flavor weak eigenstates in the three families are 
distinguished by a spontaneously broken discrete $Z_6$ chiral symmetry.
In a leading partially symmetric approximation, the model accommodates 
the family 
mass spectra, with a CKM matrix that is exactly unity.  Adding small 
asymmetries in first order perturbation theory gives a CKM matrix with 
the correct qualitative structure, with $s_{23}/s_{12}$ and $s_{13}/s_{12}$ 
naturally of order $M_s/M_b$.  
\AE
\bigskip\bigskip
\vfill\eject
\pageno=3
It has long been recognized that the hierarchical structures of the 
family mass spectra, with their large third family masses, and of the CKM 
mixing matrix, with its suppressed third family mixings, may have a common
dynamical origin.  In particular, several authors [1] have stressed 
that the 
observed pattern seems to be close to the ``rank-one'' limit, in which the 
mass matrices have the ``democratic'' form of a matrix with all matrix 
elements equal to unity, which has one eigenvalue 1 and two eigenvalues 0;  
when both up and down quark mass matrices have this form, 
they are diagonalized 
by the same unitary transformation and the CKM matrix is unity.  However, 
because the underlying dynamical basis for this choice has not been 
apparent, it has not been possible to systematically extend the rank-one   
model to one that incorporates, and relates, the observed mass and mixing 
hierarchies.  

We present in this letter a model for the quark mass and flavor mixing 
matrices, based on the underlying dynamical assumption that the three 
flavor weak eigenstates are distinguished by different eigenvalues of 
a discrete chiral $Z_6$ quantum number.  The idea that a discrete chiral 
quantum number may underlie family structure was introduced originally  
by Harari and Seiberg [2], and was developed recently by 
the author [3] in a
modified form that we follow here.  Also of relevance is the remark of  
Weinberg [4] that an unbroken discrete chiral quantum number suffices 
to enforce the masslessness of fermionic states.  Extending the general  
framework of this earlier work, we postulate that
all {\it complex} fields carry a discrete chiral family quantum number.  
Since the Higgs scalars in the standard model are complex, we introduce a 
triplet of Higgs doublets that carry $Z_6$ quantum numbers, and that 
are coupled to the fermions by Yukawa couplings 
so that the Lagrangian is $Z_6$ invariant.  
Spontaneous symmetry breaking, in which the neutral members of the three
Higgs doublets acquire vacuum expectations, then gives the 
fermion mass matrices that form the basis for our detailed analysis.  

In the simplest form of a theory with a discrete chiral $Z_6$ symmetry, 
the fundamental Lagrangian, as augmented by the instanton-induced potential, 
is invariant under the simultaneous transformations 
$$\xi_L \to \xi_L \exp(-2\pi i/6)~,~~~\xi_R \to \xi_R \exp(2\pi i/6)~~~
\eqno(1a)$$
of the unification scale fermion fields $\xi$.  
The fields in the low energy 
effective Lagrangian are in general nonlinear functionals of the fundamental 
fields.  Fermionic effective fields must be odd monomials in the fundamental 
fields, and so can come in three varieties $\psi_n$ with the discrete 
chiral transformation law
$$\psi_{nL} \to \psi_{nL} \exp(-(2n+1)2\pi i/6)~,~~~
\psi_{nR} \to \psi_{nR} \exp((2n+1)2\pi i/6)~,~~~
n=0,1,2~~~,\eqno(1b)$$
while complex bosonic effective fields must be even monomials in 
the fundamental fermion fields, and so can 
also come in three varieties $\phi_n$ 
with the discrete chiral transformation law 
$$\phi_n \to \phi_n \exp(2n 2\pi i/6)~,~~~n=0,1,2~~~.\eqno(1c)$$
Gauge fields are real fields, and since the phase in Eq.~(1c) never takes 
the value $-1$, they come in only one variety transforming with phase 
unity under discrete chiral transformations.   (Two 
varieties of gauge fields become possible when the discrete chiral group is 
$Z_{12}$ rather than $Z_6$, since $-1$ is then present in the set of even 
phases.)  Thus the minimal $Z_6$ invariant extension of the standard  
model consists of a triplicated set of fermions, and a triplicated set of 
Higgs doublets, obeying the transformation laws of Eqs.~(1b) and (1c) 
respectively, together with the usual gauge bosons,  with the Lagrangian 
constructed to be $Z_6$ invariant.  

We shall give the full Lagrangian 
structure of this theory elsewhere; 
for our present purpose it suffices to state 
that all of the usual gauge sector results are unchanged. Hence we only   
exhibit the most general Yukawa coupling of the Higgs fields to the fermions,
which grouping the the fermions $\psi_n$ into a 3-component column vector, 
takes the form 
$$\overline{Q}_L \Phi^d d_R + \overline{Q}_L \Phi^u u_R+
\overline{\psi}_L \Phi^e e_R~~~+ {\rm adjoint}.\eqno(2)$$
Here $Q_L$ and $\psi_L$ denote the usual left handed quark and lepton 
doublets, as extended into three component column vectors of doublets 
to incorporate the discrete chiral symmetry.  Similarly, $d_R$, $u_R$, 
and $e_R$ denote the usual quark and lepton right handed singlets, 
as extended 
into three component column vectors.  The $3 \times 3$ matrices $\Phi^{d,u,e}$ 

are given by 
$$\Phi^{d,e}=\sum_s P_s^{d,e} \phi_{n(s)}~,~~~
\Phi^{u}=\sum_s P_s^u \tilde \phi_{\tilde n(s)}~~~,\eqno(3a)$$
with $\phi$ and $\tilde \phi$ denoting respectively the Higgs doublet and 
the charge conjugate Higgs doublet, with $n(s)~,~ \tilde n(s)$ 
denoting the unique discrete 
chiral Higgs component, for each $s=1,2,3$, that cancels the fermionic phase 
selected by the combination matrices $P_s$. The three possibilities for $P_s$ 
are given by (with $f$ denoting $d,u,e$) 
$$P_0^{f}=\pmatrix{0       & 0       & \alpha_{13}^f    \cr
                   0       & \alpha_{22}^f   &0        \cr
                   \alpha_{31}^f &0  &0   \cr}~,~~~
P_1^{f}=\pmatrix{\alpha_{11}^f &0 &0 \cr
                 0 & 0& \alpha_{23}^f \cr
                 0 & \alpha_{32}^f & 0 \cr}~,~~~
P_2^{f}=\pmatrix{0 & \alpha_{12}^f & 0 \cr
                 \alpha_{21}^f & 0 & 0 \cr
                 0 & 0 & \alpha_{33}^f \cr}~~~,
\eqno(3b)$$
with the $\alpha$'s complex numbers.  On spontaneous symmetry breaking, 
the neutral components of each Higgs field $\phi_n$ obtains a vacuum 
expectation value $v_n$, and the Yukawa coupling of Eq.~(2) gives 
the mass term 
$$\sum_s[\overline{d}_L P_s^d v_{n(s)} d_R 
+ \overline{u}_L P_s^u v_{\tilde  n(s)} u_R
+\overline{e}_L P_s^e v_{n(s)} e_R~+ {\rm adjoint}].\eqno(4)$$
Equation (4) is the starting point for our analysis.  

We observe that 
there are two specializations of Eq.~(4) that are of interest.  The first 
is what we term the {\it partially symmetric} limit, in which the three Higgs 
expectations $v_n$ are not equal, but in which the $\alpha$'s appearing 
in each $P_s^f$ are equal to a common value $\alpha_s^f$, 
so that each Higgs couples universally to all three 
combinations of fermions that have the same overall phase under discrete 
chiral transformation.  When the partially symmetric limit is further 
specialized to the completely symmetric limit in which the products 
$\alpha_s^f v_{n(s)}$  (for $f=d,e$)  and $\alpha_s^f v_{\tilde n(s)}$ 
(for $f=u$) 
are independent of $s$, the mass matrices of Eq.~(4) for each charge 
sector become 
proportional to the ``democratic'' mass matrix with all matrix elements 
equal to unity.  

In the partially symmetric limit, the mass matrices are all proportional 
to a matrix of the general form 
$$N=\pmatrix{m_1 & m_2 & m_0\cr   
           m_2 & m_0 & m_1 \cr
           m_0 & m_1 & m_2\cr} ~~~.\eqno(5)$$
This matrix is a {\it circulant} [5] (the rows are related to one another 
by successive cyclic permutation), and thus has the property that it 
can be diagonalized by           
left and right unitary transformation matrices that are {\it independent} of 
the values of the complex numbers $m_0$, $m_1$, and $m_2$.  To show this, 
we introduce the unitary matrices $U_L$ and $U_R^{\dagger}$ given by 
$$U_L={1 \over \sqrt{3}}\pmatrix{1 & \omega^2 & \omega \cr
                                 1 & \omega & \omega^2 \cr
                                 1 & 1 & 1 \cr}~,~~~
U_R^{\dagger}={1 \over \sqrt{3}}\pmatrix{1 & 1 & 1 \cr
                               \omega^2 & \omega & 1 \cr
                               \omega & \omega^2 & 1 \cr}~~~,
                               \eqno(6a)$$
with $\omega$ the complex cube root of unity,                                
$$\eqalign{
\omega=&-{1\over 2}+{\sqrt{3}\over 2} i~,~~~
\omega^2=\omega^*= -{1\over 2}-{\sqrt{3}\over 2} i~~~~ \cr
\omega^3=&1~,~~~1+\omega+\omega^2=0~,~~~i(\omega^2-\omega)=\sqrt{3} ~~~.\cr
}\eqno(6b)$$
We then find that 
$$U_L N U_R^{\dagger}=\pmatrix{ n_1 & 0 & 0 \cr
                                0   & n_2 & 0 \cr
                                0 & 0 & n_3 \cr}~~~,\eqno(7a)$$
with the eigenvalues $n_{1,2,3}$ given by                                 
$$n_1=m_1+\omega^2 m_2+\omega m_0 ~,~~~
n_2= m_1+\omega m_2+\omega^2 m_0~,~~~
n_3=m_1+m_2+m_0~~~.\eqno(7b)$$
Because the transformation of Eq.~(6a) is independent of the values of 
the matrix elements of $N$, in the partially symmetric limit changing  
basis from weak eigenstates to mass eigenstates diagonalizes not only the 
mass term of Eq.~(4), but also the Yukawa coupling of Eq.~(2).  Hence  
in the partially symmetric limit there are no flavor changing neutral 
currents arising from tree level Higgs exchange.

To see that the magnitudes $|n_1|,|n_2|,|n_3|$ can represent arbitrary 
first, second, and third generation masses $M_1,M_2,M_3$, let us specialize 
to the case when $m_0$ is real and 
$m_2$ and $m_1$ are related by complex conjugation, 
so that $m_2^*=m_1\equiv m_R+im_I$.  
Then Eqs.~(7b) give 
$$|n_1|=|m_0-m_R+\sqrt{3}m_I|~,~~~
|n_2|=|m_0-m_R-\sqrt{3}m_I|~,~~~
|n_3|=|m_0+2m_R|~~~,\eqno(8a)$$
and we can satisfy $M_1/M_3=|n_1|/|n_3|$ and $M_2/M_3=|n_2|/|n_3|$ by taking 
$$m_R/m_0=1+{3 \over 2} {M_1+M_2\over M_3 -M_1-M_2}~,~~~
m_I/m_0={\sqrt{3}\over 2} {M_2-M_1 \over M_3-M_1-M_2}~~~.\eqno(8b)$$
Since this specialized form of $N$ can represent arbitrary mass ratios 
between the generations, so can the general form before specialization.  

Returning to the general form of N, let us determine the CKM matrix 
that results when we relax the assumption of partial symmetry, by allowing 
$P_{0,1,2}^f$ to have the general form of Eq.~(3b).  Thus, suppressing  
the superscript $f=u,d,e$ until needed, the 
mass matrix $M$ arising from Eq.~(4) now has the form
$$M=N+\epsilon~,~~~ \eqno(9a)$$
with $N$ given by Eq.~(5) and with $\epsilon$ a $3 \times 3$ complex matrix
with vanishing diagonal matrix elements (because $N$ has three independent 
parameters on its diagonal) and with arbitrary complex off-diagonal matrix 
elements. We diagonalize Eq.~(9a) in two steps.  First we apply the  
transformation of Eq.~(7a) that diagonalizes $N$, giving 
$$M^{\prime}=U_L (N+\epsilon) U_R^{\dagger}=
{\rm diag}(n_1,n_2,n_3) +\mu~,~~~\mu=U_L \epsilon U_R^{\dagger}~~~
.\eqno(9b)$$ 
Taking the inverse transformation $\epsilon=U_L^{\dagger} \mu U_R$, we 
easily find that the restrictions $\epsilon_{11}=\epsilon_{22}=\epsilon_{33}
=0$ translate into the restrictions 
$\mu_{11}=-\mu_{23}-\mu_{32}$, ~~$\mu_{22}=-\mu_{13}-\mu_{31}$,~~
$\mu_{33}=-\mu_{12}-\mu_{21}$ on the diagonal matrix elements of $\mu$.  
We now wish to find the additional 
left and right transformations $U_L^{\prime}$ and $U_R^{\prime \dagger}$  
that diagonalize $M^{\prime}$, treating 
$\mu$ as a perturbation.  Since calculation of the CKM matrix requires only 
knowledge of $U_L^{\prime}$, we first form the Hermitian matrix 
$M^{\prime}M^{\prime \dagger}$ and construct its diagonalizing unitary 
transformation.  From Eq.~(9b) we find 
$$M^{\prime}M^{\prime \dagger}={\rm diag}(|n_1|^2,|n_2|^2,|n_3|^2)+
\Delta~~~,\eqno(10a)$$
with $\Delta$ the Hermitian matrix with upper diagonal matrix elements 
given by  
$$\eqalign{
\Delta_{11}=&-2 {\rm Re}[n_1(\mu_{23}^*+\mu_{32}^*)]~,~~~
\Delta_{12}=n_1\mu_{21}^*+n_2^*\mu_{12}~,~~~
\Delta_{13}=n_1\mu_{31}^*+n_3^*\mu_{13}~~~,\cr
\Delta_{22}=&-2{\rm Re}[n_2(\mu_{13}^*+\mu_{31}^*)]~,~~~
\Delta_{23}=n_2\mu_{32}^*+n_3^*\mu_{23}~~~,\cr
\Delta_{33}=&-2{\rm Re}[n_3(\mu_{12}^*+\mu_{21}^*)]~~~.\cr
}\eqno(10b)$$
The problem of diagonalizing Eqs.(10a,~b) is textbook time-independent 
perturbation theory, and the solution is the matrix  
$$U_L^{\prime}=
\pmatrix{1 & {\Delta_{12}\over |n_1|^2-|n_2|^2}&
{\Delta_{13}\over|n_1|^2-|n_3|^2}\cr
{\Delta_{21}\over |n_2|^2-|n_1|^2} &1&{\Delta_{23}\over|n_2|^2-|n_3|^2}\cr
{\Delta_{31}\over|n_3|^2-|n_1|^2}&{\Delta_{32}\over|n_3|^2-|n_2|^2}&1\cr}~~~.
\eqno(11)$$
Restoring the superscripts $u,d$, the CKM mixing matrix is given by 
$U_{CKM}=U_L^{u \prime \dagger} U_L^{d \prime}$, which on using Eq.~(11) 
gives 
$$\eqalign{
U_{CKM}=&\pmatrix{1 & U_{12} & U_{13}\cr
                   -U_{12}^* & 1 & U_{23} \cr
                   -U_{13}^* & -U_{23}^* & 1\cr}~~~, \cr
U_{12}=&{\Delta^d_{12} \over |n_1^d|^2-|n_2^d|^2}
-{\Delta^u_{12} \over |n_1^u|^2-|n_2^u|^2}  ~~~,\cr
U_{13}=&{\Delta^d_{13} \over |n_1^d|^2-|n_3^d|^2}
-{\Delta^u_{13} \over |n_1^u|^2-|n_3^u|^2}  ~~~,\cr
U_{23}=&{\Delta^d_{23} \over |n_2^d|^2-|n_3^d|^2}
-{\Delta^u_{23} \over |n_2^u|^2-|n_3^u|^2}  ~~~.\cr
 }\eqno(12a) $$
To first order in perturbation theory, the quark masses are given by 
$$\eqalign{
M_u^2=|n_1^u|^2+\Delta_{11}^u~,~~~M_c^2=|n_2^u|^2+\Delta_{22}^u
~,~~~M_t^2=|n_3^u|^2+\Delta_{33}^u~~~,\cr
M_d^2=|n_1^d|^2+\Delta_{11}^d~,~~~M_s^2=|n_2^d|^2+\Delta_{22}^d
~,~~~M_b^2=|n_3^d|^2+\Delta_{33}^d~~~.\cr
}\eqno(12b)$$
Equations (12a~,b) are our final result for the CKM matrix, and the fermion 
masses, to first order in perturbation theory in the asymmetric 
perturbation $\mu$.

To examine the qualitative form of Eq.~(12a), let us make the approximation 
of neglecting the small quantities $M_d/M_s,M_s/M_c,M_s/M_b,M_b/M_t$, etc., 
and of neglecting the first order corrections of Eq.~(12b) to the 
masses $M_s$ and $M_b$, 
so that these are equal to $|n_2^d|$ and $|n_3^d|$ respectively.  Then 
combining the above formulas, and rephasing to put the CKM matrix in the 
standard form (to first order in the mixing angles) 
$$U_{CKM}=\pmatrix{1&s_{12}&s_{13}e^{-i\delta_{13}} \cr
                    -s_{12}&1&s_{23} \cr
                    -s_{13}e^{i\delta_{13}}& -s_{23}& 1 \cr}~~~,\eqno(13)$$
we find for the sines of the mixing angles
$$s_{12}={|\mu_{12}^d| \over M_s}~,~~~s_{13}={|\mu_{13}^d| \over M_b}
~,~~~s_{23}={|\mu_{23}^d| \over M_b }~~~,\eqno(14a)$$
and for the CP-violating phase
$$e^{-i\delta_{13}}=-{n_2^d \over |n_2^d|} {\mu_{12}^{d*} \over |\mu_{12}^d|}
{\mu_{13}^d \over |\mu_{13}^d|} {\mu_{23}^{d*}\over |\mu_{23}^d|}~~~.
\eqno(14b)$$
If the magnitudes of the off-diagonal matrix elements of $\mu$ are assumed 
roughly equal, then Eq.~(14a) tells us that the third family mixing angles 
are suppressed relative to the Cabibbo angle by a factor $M_s/M_b \sim 
1/20$, which to within factors of order unity (that can be accounted for 
by differences in the $\mu$'s) is in accord with experiment.  From the  
estimate of Eqs.~(8a~,b), we have $m_0^d \sim m_1^d \sim m_2^d\sim M_b/3$, 
with the 
fractional variation among them of order $M_s/M_b$.  Since the observed 
value of $s_{12}$ is $\sim 1/5$, we have $|\mu_{12}^d|/M_b\sim 1/100$. 
Thus the matrix elements of the initial 
partially symmetric approximation to the 
mass matrix are uniform to about five percent, and 
the additional fractional variation of the mass 
matrix elements supplied by the perturbation $\epsilon$ in Eq.~(9a), that 
breaks the partial symmetry, is of order three percent.   As a consequence,   
flavor changing neutral current effects arising from single Higgs exchange 
are suppressed by a factor of order $.03^2 \sim  10^{-3}$ relative to 
the generic estimate of Glashow and Weinberg [6], and one finds that for 
Higgs masses greater than roughly $300$ GeV, the Higgs 
exchange contribution 
to the $K_L-K_S$ mass difference is acceptably small.  

In a paper in preparation, we will give the full structure of the $Z_6$ 
invariant extension of the standard model referred to above.  Additionally,  
motivated by the model of Ref. [3], we will give a $Z_{12}$ 
invariant extension 
of the standard model that separates, under an $S(2)$ 
symmetry restriction, into two copies of the $Z_6$ model that interact 
only through the Higgs sector, and that have Yukawa couplings that are 
orthogonal linear combinations of the original $Z_{12}$ Yukawa couplings.  
Hence when the $Z_{12}$ model is close to the ``democratic'' limit, one 
of the $Z_6$ copies is also close, and will behave as discussed here; 
the second copy can have a very different (and possibly much lighter) 
mass spectrum.  In a second 
paper in preparation, we will investigate the possibilities for  
realizing a $Z_6$ or $Z_{12}$ discrete chiral symmetry structure 
in preonic models, again following the directions set out in Ref. [3].

\bigskip
\bigskip
\centerline{\bf Acknowledgments}
This work was supported in part by the Department of Energy under
Grant \#DE--FG02--90ER40542.  I wish to thank Henry Frisch, Harald Fritzsch, 
Chris Kolda, 
Jon Rosner, and Sam Treiman for stimulating conversations.  
\vfill\eject
\centerline{\bf References}
\bigskip
\noindent
\item{[1]} H. Harari, H. Haut, and J. Weyers, Phys. Lett. {\bf B78}, 459 
(1978); Y. Chikashige, G. Gelmini, R. P. Peccei, and M. 
Roncadelli, Phys. Lett.
{\bf B94}, 499 (1980); H. Fritzsch, in Proc. of Europhys. Conf. on Flavor 
Mixing in Weak Interactions, Erice, Italy (1984); C. Jarlskog, in Proc. of 
Int. Symp. on Production and Decay of Heavy Flavors, Heidelberg, Germany 
(1986); P. Kaus and S. Meshkov, Mod. Phys. Lett. {\bf A3}, 1251 (1988); 
Y. Koide, Phys. Rev. {\bf D39}, 1391 (1989); M. Tanimoto, Phys. Rev. 
{\bf D41}, 1586 (1990); G. C. Branco, J. I. Silva-Marcos, 
and M. N. Rebelo, Phys. Lett. 
{\bf B237}, 446 (1990); H. Fritzsch and J. Plankl, Phys. Lett. 
{\bf B237}, 451 (1990); H. Fritzsch, Phys. Lett. {\bf B289}, 92 (1992); 
H. Fritzsch and D. Holtmannsp\"otter, Phys. Lett. {\bf B338}, 290 (1994);
H. Fritzsch and Z. Z. Xing, Phys. Lett. {\bf B353}, 114 (1995); K. Kang 
and S. K. Kang, Phys. Rev. {\bf D56}, 1511 (1997). 
\bigskip 
\noindent
\item{[2]}  H. Harari and N. Seiberg, Phys. Lett. {\bf B102}, 263 (1981).
\bigskip
\noindent
\item{[3]}  S. Adler, hep-th/9610190 (unpublished).   
\bigskip
\noindent
\item{[4]}  S. Weinberg, Phys. Lett. {\bf B102}, 401 (1981). 
\bigskip                                                   
\noindent
\item{[5]}  M. Marcus, {\it Basic Theorems in Matrix Theory}, National 
Bureau of Standards Applied Mathematics Series no. 57 (1964), Sec. 2.13; 
H. L. Hamburger and M. E. Grimshaw, {\it Linear transformations in 
n-dimensional vector space}, Cambridge University Press (1951), pp. 94-96.  
\bigskip
\noindent
\item{[6]}  S. L. Glashow and S. Weinberg, Phys. Rev. {\bf D15}, 1958 (1977).

\vfill
\eject
\bigskip
\bye